\documentclass[aps,prb,twocolumn,superscriptaddress,longbibliography]{revtex4-2}

\usepackage{multirow}
\usepackage{amsthm,amsmath,amsfonts}
\usepackage{comment}
\usepackage{hyperref}
\usepackage{longtable,tabularx}
\usepackage{subfigure,xcolor,color}
\usepackage{siunitx}
\usepackage[utf8]{inputenc}
\usepackage[ruled,vlined]{algorithm2e}
\usepackage{graphicx}
\usepackage{dcolumn}
\usepackage{float}
\usepackage{bbm,url}
\usepackage[normalem]{ulem}
\newcommand{\msout}[1]{\text{\sout{\ensuremath{#1}}}}

\newcommand{\Tr}{\mathrm{Tr}}

\newcommand{\bra}[1]{\mbox{$\langle #1 |$}}
\newcommand{\ket}[1]{\mbox{$| #1 \rangle$}}
\newcommand{\F}{\mathcal{F}}
\newcommand{\Ho}{\mathcal{H}^{(1)}}

\newcommand{\HH}{\mathcal{H}}
\newcommand{\BH}{\mathcal{B}(\HH)}
\newcommand{\BHA}{\mathcal{B}(\HH_A)}
\newcommand{\BHB}{\mathcal{B}(\HH_B)}

\newcommand{\DD}{\mathcal{D}}
\newcommand{\Dz}{\mathcal{D}_0}
\newcommand{\Ds}{\mathcal{D}_\text{sep}}
\newcommand{\Dc}{\mathcal{D}_\text{cl}}
\newcommand{\CC}{\mathbb{C}}

\begin{document}
\author{Lexin Ding}
\affiliation{Faculty of Physics, Arnold Sommerfeld Centre for Theoretical Physics (ASC),\\Ludwig-Maximilians-Universit{\"a}t M{\"u}nchen, Theresienstr.~37, 80333 M{\"u}nchen, Germany}
\affiliation{Munich Center for Quantum Science and Technology (MCQST), Schellingstrasse 4, 80799 M{\"u}nchen, Germany}

\author{Stefan Knecht}
\email{stefan@algorithmiq.fi}
\affiliation{Algorithmiq Ltd., Kanavakatu 3C, FI-00160 Helsinki, Finland}
\affiliation{ETH Z{\"u}rich, Laboratory for Physical Chemistry, Vladimir-Prelog-Weg~2, 8093 Z{\"u}rich, Switzerland}

\author{Zolt\'an Zimbor{\'a}s}
\affiliation{Theoretical Physics Department, Wigner Research Centre for Physics, P.O.Box 49 H-1525, Budapest, Hungary}
\affiliation{Algorithmiq Ltd., Kanavakatu 3C, FI-00160 Helsinki, Finland}
\affiliation{Eötvös Loránd University, Pázmány Péter sétány. 1/C, 1117 Budapest, Hungary}

\author{Christian Schilling}
\email{c.schilling@physik.uni-muenchen.de}
\affiliation{Faculty of Physics, Arnold Sommerfeld Centre for Theoretical Physics (ASC),\\Ludwig-Maximilians-Universit{\"a}t M{\"u}nchen, Theresienstr.~37, 80333 M{\"u}nchen, Germany}
\affiliation{Munich Center for Quantum Science and Technology (MCQST), Schellingstrasse 4, 80799 M{\"u}nchen, Germany}

\title{Quantum Correlations in Molecules: From quantum resourcing to chemical bonding}

\begin{abstract}
The second quantum revolution is all about exploiting the quantum nature of atoms and molecules to execute quantum information processing tasks. To boost this growing endeavor and by anticipating the key role of quantum chemistry therein, our work establishes a framework for systematically exploring, quantifying and dissecting correlation effects in molecules. By utilizing the geometric picture of quantum states we compare --- on a unified basis and in an operationally meaningful way --- total, quantum and classical correlation and entanglement in molecular ground states. To unlock and maximize the quantum informational resourcefulness of molecules an orbital optimization scheme is developed, leading to a paradigm-shifting insight: A single covalent bond equates to the entanglement $2\ln(2)$. This novel and more versatile perspective on electronic structure suggests a generalization of valence bond theory, overcoming deficiencies of modern chemical bonding theories.
\end{abstract}

\maketitle

\section{Introduction}\label{sec:intro}

In light of the fast-approaching second quantum revolution\cite{dowling2003quantum,atzori2019second,deutsch2022second}, soon we will be able to comprehensively exploit the quantum nature of chemical systems to our advantage. This historic opportunity has boosted many studies on fermionic correlation and entanglement in the quantum information (QI) community\cite{Friis13,Friis16,Gigena15,franco2018indistinguishability,ding2020correlation,morris2020entanglement,aoto2020grassmannians,olofsson2020quantum,galler2021orbital,faba2021correlation,faba2021two,faba2022Lipkin,benatti2020entanglement,sperling2022entanglement}, with an emphasis on the conceptual formalism and resource utilization aspects. Independently, fermionic correlation has been used in the quantum chemistry (QC) community to describe the electronic structure of chemical systems \cite{Reiher12,Reiher13,Reiher14,Reiher15,Ayers15a,Ayers15b,Bogus15,Szalay17,Reiher17b,Legeza18,Legeza19,pusuluk2021classical} and to improve as well as optimize the initial ansatz of numerical methods\cite{Eisert16mode,Reiher16,turner2017optimal}. Yet, only rudimentary QI tools have thus far been widely adopted for such type of studies. It is therefore the ultimate motivation of our work to propel the on-going second quantum revolution by combining the expertise of both the QI and QC community and to harness synergies. To achieve this,
three important facets need to be furnished. Firstly, the dissection of classical and quantum correlation effects still needs to be distinctly and visibly advocated, evidenced by the fact that only the von Neumann entropy has been so far systematically applied. Secondly, this simple quantity is even erroneously calculated, by ignoring important fermionic superselection rules\cite{wick1970superselection,wick1997intrinsic} (SSR). Violation of these rules, in the context of the second quantum revolution, leads to a gross overestimation of the accessible quantum resource. Thirdly, the subject of correlation analysis has been primarily limited to the canonical, delocalized orbitals. Accordingly, these three points urge us to provide in the present work tailored tools for operationally meaningful quantification and categorization of the correlation in chemical system.

To this end, we turn to the plethora of correlation quantities offered by the QI community, although their application to chemical systems is not without hurdles. Relevant chemical concepts and interesting phenomena are often concerned with a particular subregion of molecular systems. For instance, the concept of a chemical bond often refers to just two neighboring atoms. Yet, their quantum state is not pure anymore due to their interaction with  the other atoms in the molecule. Accordingly, the easy-to-calculate von Neumann entropy is not a valid measure of correlation or entanglement anymore. Moreover, various definitions of quantum correlation, classical correlation\cite{ollivier2001quantum,luo2010geometric,henderson2001classical} and entanglement\cite{vedral1997quantifying,wootters1998entanglement,Horo09} for mixed states from QI have often different conceptual origins. The unpleasant consequence is that one cannot compare the respective values against each other, just as it does not make sense in chemistry to compare molecular energy values referring to different systems of physical units. Hence, it is one of our main goals to offer an appealing geometric picture of quantum states in which all these correlation quantities can be defined on the same footing.

With a suitable QI framework at hand we can then address the following fundamental question: What is the QI resource content of a molecule in general and of a chemical bond in particular?
Based on an orbital optimization scheme, we identify fully localized orbitals as the adequate reference basis for realizing and extracting this total entanglement. Due to the connection between electronic structure and entanglement, this key observation of our work
will also bridge the two pillars of chemical bonding theories: molecular orbital\cite{muller1994glossary,roos2007reaching} (MO) theory and valence bond\cite{shaik2004valence,goddard1973generalized} (VB) theory. In MO theory, a bond of order one is represented by a fully occupied bonding molecular orbital (and its empty antibonding partner). By contrast, VB theory describes a bond with a pair of coupled atomic orbitals, thus offering an attractive local perspective of bonding. Adversely, VB theory by itself does not reach modern chemical accuracy, and a quantitative measure for the correlation between the atomic orbitals is lacking. Our work fuses these two distinctively different pictures, and makes use of the correlation between atomic-like molecular orbitals to unveil the unique orbital pairing structures in chemical bonds: A single covalent bond is equivalent to the orbital entanglement content $2\ln(2)$. This value exceeds by an order of magnitude the numbers reported in previous studies  \cite{Reiher13,Reiher14,Reiher15,Ayers15a,Ayers15b,Bogus15,Szalay17,Reiher17b,Legeza18,Legeza19}. This in turn emphasizes that QI tools applied to delocalized orbitals describe primarily the validity of the independent electron-pair picture rather than the bonding structure of molecular systems.

The paper is structured as follows. We start with a brief overview of fermionic correlation in Section \ref{sec:QIT} for the less experienced audience, and for later use. Then in Section \ref{sec:analytic} we illustrate these concepts with two analytic examples, and explore the relation between orbital entanglement and bonding. Finally in Section \ref{sec:comput} and \ref{sec:results} we demonstrate in realistic molecular systems the effect of orbital localization on maximizing their quantum resource, as well as revealing their bonding structures.

\section{Concepts}\label{sec:QIT}

In this Section we will give a brief overview of fermionic correlation and entanglement. Specifically, we will explain 1) how are various correlation quantities such as entanglement and quantum correlation quantified in QI, and 2) how can one transfer these QI concepts, which are formulated for distinguishable systems, to indistinguishable fermions. Both aspects are not only crucial for understanding the main results of this work, but also potentially beneficial for specialists from QI or QC communities who would like to also work on this interdisciplinary topic. Experts on fermionic entanglement shall feel  free to skip this section and return in case any unfamiliar concepts may appear in the proceeding Sections.

\subsection{Geometry of quantum states}\label{subsec:geom}
In order to set the stage for what follows, it will be essential to first recall and discuss basic geometric aspects of the space of quantum states. It is exactly this geometric picture which will namely allow us to quantify and compare on a unified basis total, quantum and classical correlation and entanglement in molecular quantum systems.

\begin{figure}[t]
    \centering
    \includegraphics[scale=0.3]{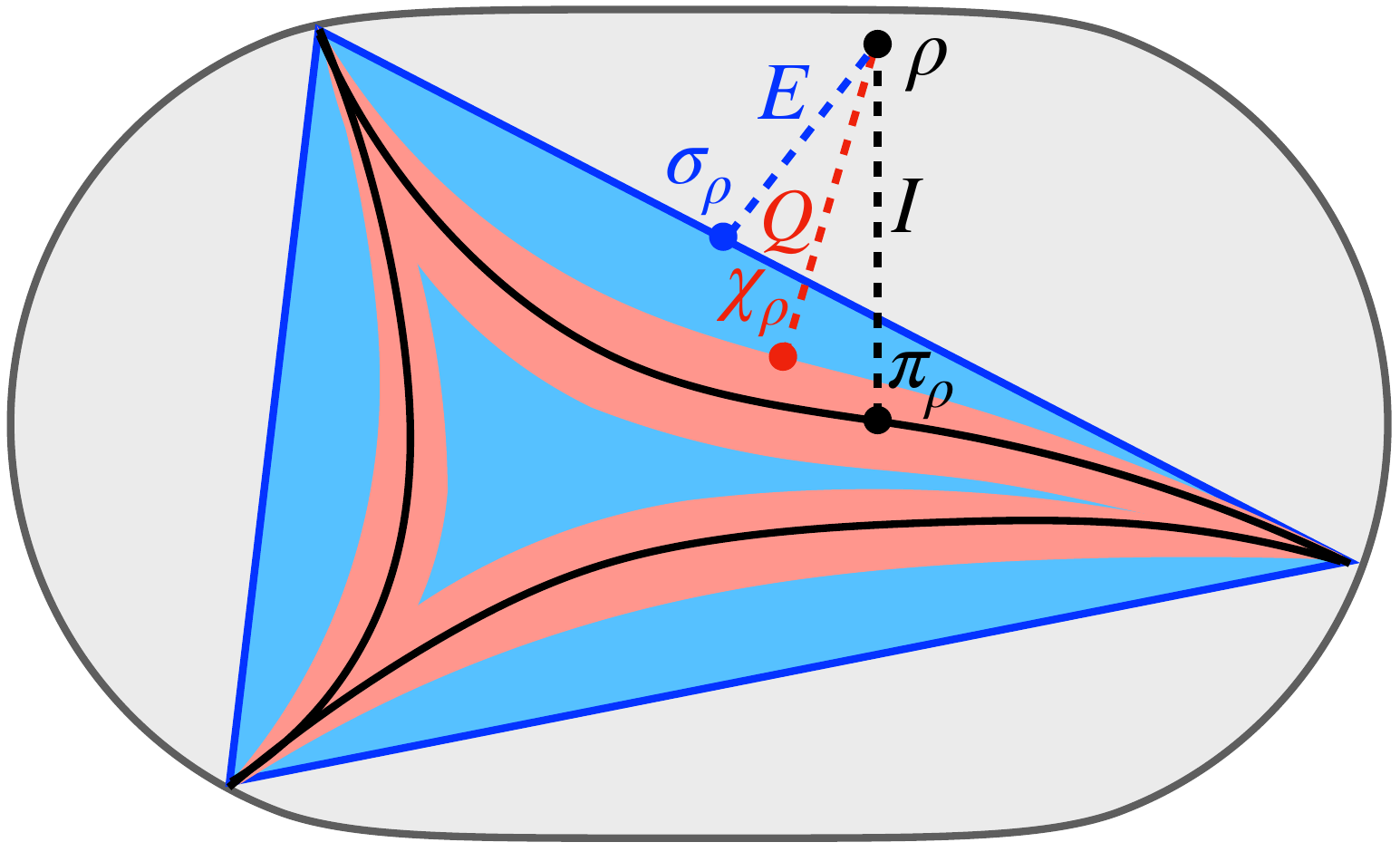}
    \caption{Schematic illustration of the state space $\mathcal{D}$. The subsets of uncorrelated ($\Dz$), classically correlated ($\Dc$) and separable states ($\Ds$) are shown in black, red and blue, respectively. The corresponding measures of total correlation $I$, quantum correlation $Q$ and entanglement $E$ are given by the quantum relative entropy \eqref{relEnt} of $\rho$ minimized with respect to those sets, with corresponding minimizers $\pi_{\rho}$, $\chi_{\rho}$ and $\sigma_{\rho}$.}
    \label{fig:states}
\end{figure}

Let us start by considering a complex finite-dimensional Hilbert space $\HH$ of dimension $d$ and denote the algebra of linear operators acting on $\HH$ by $\BH$. The corresponding set $\DD$ of density operators is given by all Hermitian operators $\rho$ on $\HH$ which are positive-semidefinite $\rho \geq 0$ (i.e., $\rho$ has non-negative eigenvalues), and trace-normalized to unity,
\begin{equation}\label{D}
\DD=\{\rho\in \BH\,|\, \rho^\dagger =\rho,\, \rho \geq 0, \, \Tr[\rho]=1\}\,.
\end{equation}
As it is illustrated in Figure \ref{fig:states}, the set $\DD$  is convex since the convex combination $p \rho + (1-p)\tilde{\rho}$ of any two density operators $\rho,\tilde{\rho}\in \DD$ and any $0\leq p \leq 1$ is again a density operator.
In order to develop a better intuition for $\DD$, we observe that a density operator $\rho$ lies on the boundary of $\DD$ if it is not strictly positive, that is, at least one of its eigenvalues vanishes. A particularly important subset of boundary points is given by the extremal points of $\DD$. These are per definition those ``points'' $\rho \in \DD$ which cannot be written as convex combinations of other points in $\DD$. One easily verifies that a density operator $\rho$ is extremal if and only if it is a pure state
\begin{equation}
    \rho \equiv \ket{\Psi}\!\bra{\Psi}\quad \text{with}\  \ket{\Psi} \in \HH\ ,
    \label{eqn:rho_pure_state}
\end{equation}
or, equivalently, if $\rho$ has eigenvalues $\{1,0,\ldots,0\}$.

Since the space $\DD$ of density operators is a subset of the Euclidean vector space of linear operators on $\HH$ (or equivalently just $\CC^{d \times d}$) we can introduce a notion of distances and angles in a straightforward manner. To this end, we introduce the Hilbert-Schmidt inner product on $\BH$,
\begin{equation}\label{HS}
\langle \hat{A}, \hat{B} \rangle \equiv \Tr[\hat{A}^\dagger \hat{B}]
\end{equation}
where $\hat{A}, \hat{B} \in \BH$ are linear operators.
By employing either the induced norm, any other metric or a generalized distance function we can then quantify the similarity of quantum states. The huge advantage of this approach lies in the universality of its predictions: Whenever two density operators are close to each other, the same follows as a mathematical consequence for their expectation values for \emph{any} choice of observable. A prominent  generalized distance function\footnote{In a strict mathematical sense the quantum relative entropy does not define a distance function. For instance, it is not symmetric, i.e., $S(\rho||\pi) \neq S(\pi||\rho)$ and it does in general not obey the triangle inequality.} is given by the quantum relative entropy\cite{Lindblad74,vedral1997quantifying}
\begin{equation}\label{relEnt}
S(\rho||\sigma) \equiv  \Tr[\rho(\ln{\rho}-\ln\sigma)]\,.
\end{equation}
Its relevance for quantum sciences in general and our work in particular originates from  its distinctive information-theoretical meaning. It describes by concise means \emph{``how difficult it is to distinguish the state $\rho$ from the state $\sigma$''} \cite{vedral2002role} (see also Ref.~\cite{hiai1991proper}).

\subsection{Overview of various correlation types}\label{subsec:Corrtypes}
The quantum information theoretical concepts of correlation and entanglement refer to a notion of subsystems \cite{werner1989quantum,nielsen2002quantum,Horo09}. In order to discuss them in the  context of \textit{bipartite} systems, we assume that our total system comprises two \emph{distinguishable} subsystems $A$ and $B$. The corresponding Hilbert space then takes the form
\begin{equation}\label{HilbertAB}
\HH = \HH_A \otimes \HH_B
\end{equation}
and likewise for the algebra of observables,
\begin{equation}\label{algebraAB}
\BH = \BHA \otimes \BHB\,.
\end{equation}
To motivate the concept of total correlation let us consider two local measurements, with corresponding observables $\hat{A} \in \BHA$ of subsystem A and $\hat{B} \in \BHB$ of subsystem B. The correlation between these two measurements is described by the correlation function
\begin{eqnarray}\label{corrfuncAB}
\mathcal{C}(\hat{A},\hat{B}) &\equiv& \langle \hat{A}\otimes \hat{B}\rangle_{\rho} - \langle \hat{A}\otimes \hat{1}_B\rangle_{\rho} \langle \hat{1}_A\otimes \hat{B}\rangle_{\rho} \nonumber  \\
&\equiv &\langle \hat{A}\otimes \hat{B}\rangle_{\rho} - \langle \hat{A}\rangle_{\rho_A} \langle \hat{B}\rangle_{\rho_B}\,,
\end{eqnarray}
where $\rho_{A/B} \equiv \Tr_{B/A}[\rho_{AB}]$ denotes the reduced density operator of $\rho$ for subsystem $A/B$ and $\hat{1}_{A/B}$ the identity operator on $\HH_{A/B}$.
The crucial observation is now that a vanishing correlation function, $\mathcal{C}(\hat{A}, \hat{B})=0$, does not necessarily imply the same for any other pair of local observables $\hat{A}', \hat{B}'$. This in turn strongly suggests the notion of uncorrelated states: A density operator $\rho$ is called uncorrelated  if and only if its correlation function \eqref{corrfuncAB} vanishes for \emph{all} pairs of local observables $\hat{A}, \hat{B}$. As long as the algebra $\mathcal{A}_{A/B}$ of physical observables of system $A/B$ includes all linear operators on  $\HH_{A/B}$, this is equivalent to the factorization of $\rho$ into its reduced states, \mbox{$\rho = \rho_A \otimes \rho_B$}.
The corresponding set
\begin{equation}
    \Dz \equiv \{\rho = \rho_A\otimes \rho_B\}
\end{equation}
of all uncorrelated states is schematically illustrated in Figure \ref{fig:states}. By referring to this geometric picture, a straightforward definition of the \textit{total} correlation $I(\rho)$ contained in a quantum state $\rho$ follows. It is given as the minimal ``distance'' of $\rho$ to the set $\Dz$,
\begin{eqnarray}\label{eqn:I}
I(\rho) &\equiv& \min_{\pi \in \Dz} S(\rho||\pi) \nonumber \\
&=& S(\rho_A)+ S(\rho_B)-S(\rho)\,,
\end{eqnarray}
measured in terms of the quantum relative entropy $ S(\cdot||\cdot)$, where $S(\rho)\equiv -\Tr[\rho \ln{\rho}]$ denotes the von Neumann entropy. Remarkably, the minimization in \eqref{eqn:I} can be executed analytically, leading to the mutual information (second line) where the closest uncorrelated state follows as $\pi_{\rho}=\rho_A \otimes \rho_B$\cite{modi2010unified}. Coming back to our motivation, we present an important relation between the total correlation and correlation functions which follows directly from  results presented in \cite{wolf2008area,Watrous} (see Ref.~\cite{CS21lecture} for a detailed derivation),
\begin{equation}\label{CorrFversusI}
\frac{|\mathcal{C}(\hat{A},\hat{B})|}{\|\hat{A}\|_\text{op}\,\|\hat{B}\|_\text{op}} \leq  \sqrt{2}\, \sqrt{I(\rho)}\,.
\end{equation}
Here, $\|\hat{A}\|_\text{op}$ denotes the operator norm of $\hat{A}$, i.e., the largest absolute value of its eigenvalues.
Relation \eqref{CorrFversusI} confirms in quantitative terms that whenever a quantum state is close to the set $\Dz$ of uncorrelated states then its correlation  function is small for \emph{any} choice of local observables $\hat{A},\hat{B}$. This highlights again the strength of quantum information theoretical concepts which is based on the universal character of their predictions.

Due to the information theoretical meaning of the quantum relative entropy \eqref{relEnt}, the total correlation \eqref{eqn:I} quantifies the additional information content in the state $\rho$ beyond the information content in $\rho_A \otimes \rho_B$ (local information). The term ``total'' emphasizes here that $I(\rho)$ includes both classical and quantum correlations. In order to explore and conclusively understand the significance of either correlation part in chemical bonding and quantum chemistry in general, concise definitions of quantum correlation and classical correlations are needed as well. We first start, however, with the most prominent type of quantum correlation, the entanglement.

Separable or unentangled states are precisely those states that can be prepared by distant laboratories using local operations and classical communication (LOCC) only\cite{werner1989quantum}. With local operations, two distant parties can prepare any uncorrelated state $\rho_A\otimes\rho_B$. In combination with classical communication arbitrary mixtures of uncorrelated states can be realized. Hence, the convex set of separable states is given by
\begin{equation}\label{eqn:Ds}
\begin{split}
&\mathcal{D}_\text{sep} \equiv \left\{  \sigma \!=\! \sum_{i} p_{i} \sigma_A^{(i)} \otimes \sigma_B^{(i)}, p_{i}\! >\! 0,\sum_{i} p_{i}\!=\!1 \right\}.
\end{split}
\end{equation}
As it is illustrated in Figure \ref{fig:states}, this set is indeed nothing else than the convex hull of $\Dz$.
Any state $\rho$ that is not separable is called entangled. In complete analogy to the quantification of total correlation, the entanglement in $\rho$ is quantified through the geometric picture\cite{henderson2000information}
\begin{equation}\label{eqn:E}
E(\rho) \equiv \min_{\sigma \in \Ds} S(\rho||\sigma) = S(\rho||\sigma_\rho)\,.
\end{equation}
For general mixed states, no closed expression exists for this \emph{relative entropy of entanglement}, except for highly symmetric states\cite{ding2022quantifying}. This unpleasant fact is due to the involved structure of the set \eqref{eqn:Ds} of separable states and the resulting complexity of its boundary.
For pure states $\rho\equiv \ket{\Psi}\!\bra{\Psi}$, however, \eqref{eqn:E} simplifies to a closed expression\cite{vedral1997quantifying}
\begin{eqnarray} \label{eqn:Epure}
E(|\Psi\rangle\!\bra{\Psi}) = S(\rho_A) = S(\rho_B).
\end{eqnarray}
As a consistency check, we recall that the spectra of the reduced states $\rho_A$ and $\rho_B$ are indeed identical, as it is guaranteed by the Schmidt decomposition\cite{ekert1995entangled} of $|\Psi\rangle$.

Entanglement is certainly a key concept of quantum physics\cite{EisertReview,kitaev2006topological,Laflor16} and its broad significance as a resource for realizing quantum information processing tasks is undeniable\cite{wootters1998quantum,bouwmeester1997experimental,ursin2007entanglement,jennewein2000quantum}. Yet, there also exist quantum correlations \emph{beyond} entanglement. In order to explain this crucial aspect of our work, and in analogy to the definition of total correlation and entanglement, we first characterize the family of states with zero quantum correlation\cite{oppenheim2002thermal,luo2008classical,modi2010unified} (illustrated as pink region in Figure \ref{fig:states}):
\begin{equation} \label{eqn:Dcl}
    \Dc \equiv \Big\{\,\chi\!=\!\sum_{i,j}p_{ij} |i\rangle\!\langle i|\!\otimes\! |j\rangle\!\langle j| \,\Big\}.
\end{equation}
On the right-hand side, $\{\ket{i}\}$ and $\{\ket{j}\}$ could be \emph{any} sets of orthonormal states in the Hilbert spaces of subsystems $A$ and $B$, respectively and $p_{ij}>0$, $\sum_{ij}p_{ij}=1$. The states in \eqref{eqn:Dcl} are indeed \emph{classical} in the following sense.  There exists joint local measurement $\{P_A^{(i)}\otimes P_B^{(j)}\}$ which leave the state $\rho$ unchanged, namely those with $\{P_{A/B}^{(i)}\}$ projecting onto the local eigenstates $\{\ket{i}\}$ and $\{\ket{j}\}$ of $\rho$. Therefore the correlation encoded in the resulting joint probability distribution $\{p_{ij}\}$ has to be purely classical\cite{luo2008classical}.
Any state not in $\Dc$ then contains quantum correlation.

By referring again to the geometric picture of quantum states, the quantum correlation in $\rho$ is quantified as its minimized quantum relative entropy with respect to the set of classically correlated states\cite{modi2010unified} (with the minimizer denoted by $\chi_\rho$).
\begin{equation}
Q(\rho) \equiv \min_{\chi \in \Dc} S(\rho||\chi) \equiv S(\rho||\chi_\rho).\label{eqn:Q}
\end{equation}

Note that $\Dz \subseteq \Dc$ since every uncorrelated state \mbox{$\rho_A \otimes \rho_B \equiv \big(\sum_i p_A^{(i)} \ket{i}\!\bra{i}\big) \otimes \big(\sum_j p_B^{(j)} \ket{j}\!\bra{j}\big)$} can be written as in \eqref{eqn:Dcl}, namely with $p_{ij} = p_A^{(i)} p_B^{(j)}$. On the other hand, the set $\Ds$ in \eqref{eqn:Ds} is strictly larger than $\Dc$. This is due to the fact that in the former $\{\sigma_{A/B}^{i}\}$ are typically not  simultaneously diagonalizable for all $i$. Comparing \eqref{eqn:I}, \eqref{eqn:E} and \eqref{eqn:Q} we then get the following instructive inclusion relations
\begin{equation}\label{eqn:relset}
\DD_0 \subseteq \DD_\text{cl} \subseteq \DD_\text{sep}.
\end{equation}
Thanks to the underlying geometric picture --- which provides a unified basis for quantifying the different correlation types --- this can directly be translated into relations between the respective measures
\begin{equation}\label{eqn:relmeas}
I(\rho) \geq Q(\rho) \geq E(\rho).
\end{equation}

Finally, we present the classical counterpart of \eqref{eqn:Q}, the classical correlation. To motivate its measure we first rewrite \eqref{eqn:Q} as\cite{luo2008using}
\begin{equation}
\begin{split}
&Q(\rho) 
\\
&\quad= \min_{\{P_A^{(i)}\},\{P_B^{(j)}\}} S(\rho\|\sum_{ij} P_A^{(i)}\otimes P_B^{(j)} \rho P_A^{(i)} \otimes P_B^{(j)})
\end{split}
\end{equation}
where $\{P_A^{(i)}\}$ and $\{P_B^{(j)}\}$ are two projective measurements, satisfying $\big(P_{A/B}^{(i)}\big)^2 = P_{A/B}^{(i)}$ and $\sum_i P_{A/B}^{(i)} = \openone_{A/B}$. The closest classical state $\chi_\rho$ is then the state resulting from $\rho$ after the optimal projective measurements has been performed. Accordingly, the total correlation in $\chi_\rho$ is then nothing else than the classical correlation in $\rho$\cite{modi2010unified}
\begin{equation}\label{eqn:C}
C(\rho) \equiv I(\chi_{\rho}).
\end{equation}
Since quantum states cannot be dissected into classical and quantum parts in a strict mathematical sense, it is not surprising that our measures do typically not obey the relation $I = Q + C$. However, this exact additive relation is valid whenever the closest classical state $\chi_\rho$ and the closest uncorrelated state $\pi_\rho$ have the same eigenstates. For a simple proof of this statement see Appendix \ref{app:corr_sum}.

\subsection{Application to systems of electrons}\label{subsec:fermions}

In this section, we will discuss how the previously defined quantum information concepts are adopted to electronic structure theory. For this, we need to establish in a first step a notion of subsystems in fermionic quantum systems and also learn how to deal with related conceptual peculiarities. All these aspects are absolutely vital for establishing a quantum information framework for
quantifying in an operationally meaningful way total, quantum and classical correlation and entanglement in molecular ground states.

We start by assigning a reference basis to the one-particle Hilbert space $\HH^{(1)} = \HH_l^{(1)} \otimes \HH_s^{(1)}$, by referring to a set of orthonormal spin-orbitals $\{|\chi_i \rangle \otimes |\sigma\rangle\}_{i=1}^D$, with $|\chi_i\rangle \in \HH_l^{(1)}$ being the orbital state and $\sigma \in \{\uparrow,\downarrow\}$ describing the spin degree of freedom. The corresponding Fock space $\mathcal{F}[\HH^{(1)}]$ is then spanned by the configuration states
\begin{equation}\label{eqn:config}
    |n_{1\uparrow},n_{1\downarrow}, \ldots,n_{D\downarrow} \rangle \equiv \big(f^\dagger_{\chi_1\uparrow}\big)^{n_{1\uparrow}}\ldots \big(f^\dagger_{\chi_D\downarrow}\big)^{n_{D\downarrow}}|\Omega\rangle\,.
\end{equation}
Here, $f^{(\dagger)}_{\chi_i\sigma}$ denotes the annihilation (creation) operator associated to the spin-orbital $|\chi_i\rangle\otimes|\sigma\rangle$. They satisfy the fermionic anti-commutation relations $\{f^{(\dagger)}_{\chi_i \sigma},f^{(\dagger)}_{\chi_j \sigma'}\} = 0$ and $\{f^{\dagger}_{\chi_i \sigma},f^{\phantom{\dagger}}_{\chi_j \sigma'}\} = \delta_{i,j} \delta_{\sigma,\sigma'} \openone$, with $\{A,B\}\equiv AB + BA$, and $|\Omega\rangle$ is the vacuum state. With respect to the ordered basis $\{|\chi_i \rangle \otimes |\sigma\rangle\}_{i=1}^D$, we now establish the notion of subsystems. For the purpose of this paper, we focus primarily on the partitioning of orbitals $\{|\chi_i \rangle\}$ into two subsets $A$ and $B$. This means effectively to divide the orbital one-particle Hilbert space into two complementary subspaces of dimensions $D_{A/B}$, $\HH_l^{(1)} = \HH^{(1)}_{l,A} \oplus \HH^{(1)}_{l,B}$
and thus $\HH^{(1)} = \HH^{(1)}_A \oplus \HH^{(1)}_B$ where $\HH^{(1)}_{A/B} \equiv \HH^{(1)}_{l,A/B} \otimes \HH_s^{(1)}$. This splitting in turn induces a tensor-product decomposition on the Fock state,
\begin{equation}\label{eqn:FockAB}
\F[\Ho_A \oplus \Ho_B] \cong \F[\Ho_A] \otimes \F[\Ho_B],
\end{equation}
through the map
\begin{eqnarray}\label{configsplit}
&&\ket{n_{1\uparrow}, n_{1\downarrow}, \ldots,n_{D_A \downarrow},n_{D_{A+1} \uparrow},\ldots,n_{D \downarrow}} \mapsto \nonumber \\
&&\qquad  \ket{n_{1\uparrow}, n_{1 \downarrow},\ldots,n_{D_A \downarrow}} \!\otimes\!  \ket{n_{D_{A+1}  \uparrow}, ,\ldots,n_{D \downarrow}}.
\end{eqnarray}

It is important to note, however, that such a tensor-product decomposition does not hold on the level of fermionic operators that are defined within the respective subsystems. This is clear from the observation that the creation and annihilation operators associated with spin-orbitals in subsystem $A$ and $B$ do not commute with each other, and, as a result, cannot be considered local observable operators. The immediate consequences is the violation of special relativity, exemplified by the possibility of superluminal signaling\cite{johansson2016comment,ding2020concept}. We resolve this by invoking the fermionic parity superselection rule\cite{wick1970superselection,SSR} (P-SSR). The P-SSR excludes observables that do not commute with the local particle number parity operator $\hat{\mathcal{P}}^{(A/B)} = \hat{P}^{(A/B)}_\text{even} - \hat{P}^{(A/B)}_\text{odd}$, where $\hat{P}^{A/B}_\tau$ is the projection onto the $\tau \in \{\text{even,odd}\}$ parity subspace acting on subsystem $A/B$. As a result, the accessible correlation and entanglement in a bipartite state $\rho_{AB}$ is reduced to those in the superselected state\cite{bartlett2003entanglement,ding2020concept}
\begin{equation}
    \rho_{AB}^\text{P} = \sum_{\tau, \tau' = \text{even, odd}} \hat{P}_\tau^{(A)} \otimes \hat{P}^{(B)}_{\tau'} \rho_{AB} \hat{P}^{(A)}_\tau \otimes \hat{P}^{(B)}_{\tau'},
\end{equation}
since, from an operational point of view, $\rho_{AB}$ and $\rho_{AB}^{\text{P}}$ are equivalent. To summarize, the correlation quantities $X = I, C, Q, E$ defined in Section \ref{subsec:Corrtypes} under P-SSR can be calculated as
\begin{equation} \label{eqn:SSRC}
    X^\text{P}(\rho_{AB}) = X(\rho^\text{P}_{AB}).
\end{equation}
When the creation or annihilation of \textit{pairs} of particles is also not possible, this results in an even more  restrictive particle number superselection rules\cite{wick1970superselection,verstraete2003quantum} (N-SSR). In this case we simply replace $\rho^{P}_{AB}$ in \eqref{eqn:SSRC} with the N-SSR superselected state
\begin{equation}
    \rho_{AB}^\text{N} = \sum_{m=0}^{2D_A} \sum_{n=0}^{2D_B}  P_m^{(A)} \otimes P^{(B)}_{n} \rho_{AB} P^{(A)}_m \otimes P_{n}^{(B)},
\end{equation}
where $P_m^{(A/B)}$ is the projection onto the $m$-particle subspace acting on subsystem $A/B$.

We remark that if one is only interested in the numerical structure of the quantum state, it is possible to quantify correlation and entanglement without superselection rules, simply by mapping the fermionic state to that of a spin system via the respective  Jordan-Wigner transformation. By contrast, if the orbital entanglement in the molecules is to be accessed or utilized, e.g. through an entanglement swapping protocol from molecule to quantum registers, the inclusion of superselection rules is operationally crucial. Accessing orbital entanglement requires measurement or more generally operations to be performed on the orbitals, which are limited precisely by the superselection rules. Although at the moment, perfect control over every element within a molecule or arbitrary operations on orbitals is not yet possible, this status may greatly improve soon with the on-going second quantum revolution\cite{krylov2020orbitals}. If superselection rules are ignored, one would grossly overestimate the accessible correlation and entanglement.

Finally, we introduce the primary objects of interests, namely the {one}- and two-\textit{orbital} reduced density matrices (RDMs). Formally, the one- and two-orbital RDMs of a pure state $|\Psi\rangle$ are defined via the following requirements
\begin{subequations} \label{eqn:ordm}
\begin{align}
    \rho_i &: \Tr[\rho_i \hat{\mathcal{O}}] = \langle \Psi |\hat{\mathcal{O}} |\Psi\rangle, \quad \forall \hat{\mathcal{O}} \in \mathcal{A}_i \label{eqn:1ordm}
    \\
    \rho_{i,j} &: \Tr[\rho_{ij} \hat{O}] = \langle \Psi |\hat{\mathcal{O}} |\Psi\rangle, \quad \forall \hat{\mathcal{O}} \in \mathcal{A}_{ij}. \label{eqn:2ordm}
\end{align}
\end{subequations}
In practice, they are computed from partial two- and four-\textit{particle} RDMs\cite{amosov2017spectral}, respectively. Exploiting the symmetry of the overall quantum state reduces further the computational cost\cite{Bogus15}. For example, it is common to restrict the calculation of the molecular ground state to a predefined, fixed particle number $N$ and spin magnetization $m_S$\ (for a given spin state $2S+1$). Consequently, the one- and two-orbital RDM can only be mixtures of fixed electron-number and magnetization states. In other words, the one-orbital RDM is diagonal in the fixed particle number and magnetization basis
\begin{equation}
\begin{split}
&\quad \rho_i =
\\
&\begin{pmatrix}
\langle f^{\phantom{\dagger}}_{i\uparrow}\!f^\dagger_{i\uparrow}\!f^{\phantom{\dagger}}_{i\downarrow}\!f^\dagger_{i\downarrow} \rangle & 0 & 0 & 0
\\
0 & \langle f^\dagger_{i\uparrow}\!f^{\phantom{\dagger}}_{i\uparrow}\!f^{\phantom{\dagger}}_{i\downarrow}\!f^\dagger_{i\downarrow} \rangle & 0 & 0
\\
0 & 0 & \langle f^{\phantom{\dagger}}_{i\uparrow}\!f^\dagger_{i\uparrow}\!f^\dagger_{i\downarrow}\!f^{\phantom{\dagger}}_{i\downarrow} \rangle & 0
\\
0 & 0 & 0 & \langle f^\dagger_{i\uparrow}\!f^{\phantom{\dagger}}_{i\uparrow}\!f^\dagger_{i\downarrow}\!f^{\phantom{\dagger}}_{i\downarrow} \rangle
\end{pmatrix}
\end{split}
\end{equation}
and the two-orbital RDM is block diagonal, the form of which we refer to Ref.~\cite{Bogus15}.

\section{Analytical examples}\label{sec:analytic}

In this section we demonstrate with two analytic examples (i) the strong influence of superselection rules on the accessible correlation and entanglement and (ii) the subtle connection between entanglement and chemical bonding.

\subsection{Single Electron State}\label{subsec:1el}

We consider a single polarized electron within the manifold of two orbitals $A$ and $B$. The state of the electron is then simply a superposition of the form
\begin{equation} \label{eqn:single}
    |\Psi(\theta,\varphi)\rangle = \cos(\theta)|1_A,0_B\rangle + e^{i\varphi}\sin(\theta)|0_A,1_B\rangle,
\end{equation}
where $|n_A,n_B\rangle \equiv (f^\dagger_A)^{n_A}(f^\dagger_B)^{n_B}|\Omega\rangle$ are local occupation number eigenstates, and $\theta \in [0,\pi)$, $\varphi \in [0,2\pi)$. Such a state belongs to the one-particle Hilbert space $\mathcal{H}^{(1)}$ isomorphic to that of a single qubit, which is a subspace of the four-dimensional total Fock space $\mathcal{F}[\mathcal{H}^{(1)}]$. Referring to the tensor product between the two local Fock spaces $\mathcal{F}[\mathcal{H}^{(1)}]=\mathcal{F}[\mathcal{H}^{(1)}_A] \!\otimes\! \mathcal{F}[\mathcal{H}^{(1)}_B]$, the density operator $\rho(\theta,\varphi)$\ is given in the basis $|0_A,0_B\rangle,|1_A,0_B\rangle,|0_A,1_B\rangle,|1_A,1_B\rangle$ by,
\begin{equation}
    \begin{split}
        \rho(\theta,\varphi) &= |\Psi(\theta,\varphi)\rangle\langle\Psi(\theta,\varphi)|
        \\
        &= \begin{pmatrix}
        0 & 0 & 0 & 0
        \\
        0 & \cos^2(\theta) & \frac{e^{i\varphi}}{2}\sin(2\theta) & 0
        \\
        0 & \frac{e^{-i\varphi}}{2}\sin(2\theta) & \sin^2(\theta) & 0
        \\
        0 & 0 & 0 & 0
        \end{pmatrix} .
    \end{split}
\end{equation}
Since $\rho(\theta,\varphi)$ is pure (cf.~Eq.~\eqref{eqn:rho_pure_state}), the associated entanglement $E$\ and quantum correlation $Q$\ are the same, and so are the closest separable $\chi_\rho$ and classical states $\sigma_\rho$, respectively. The closest product state $\pi_\rho$\ as well as the classical and separable states (in this case they coincide) $\chi_\rho$ to $\rho(\theta,\varphi)$ is diagonal in this basis with
\begin{equation}
    \begin{split}
        \mathrm{diag}(\pi_\rho) &= (\cos^4(\theta),\frac{1}{4}\sin^4(2\theta),\frac{1}{4}\sin^4(2\theta), \sin^4(\theta)),
        \\
        \mathrm{diag}(\chi_\rho) &= (0,\cos^2(\theta),\sin^2(\theta),0).
    \end{split}
\end{equation}
 Moreover, the total correlation $I$, quantum correlation $Q$, classical correlation $C$, and entanglement $E$ of  $\rho(\theta,\varphi)$ are given by
\begin{equation}
    \begin{split}
        \frac{1}{2} I(\rho) &= Q(\rho) = C(\rho) = E(\rho)
        \\
        &= -[\cos^2(\theta) \ln(\cos^2(\theta)) \!+\!\sin^2(\theta) \ln(\sin^2(\theta))]
        \\
        &\equiv P(\theta).
    \end{split}
\end{equation}
In the presence of a superselection rule, the superselected state loses all coherence between different local particle number sectors
\begin{equation}
\begin{split}
    \rho^\text{P, N}(\theta,\varphi) &= \cos^2(\theta)|1_A,0_B\rangle\langle1_A,0_B|
    \\
    &\quad+\sin^2(\theta)|0_A,1_B\rangle\langle0_A,1_B|\ .
\end{split}
\label{eq:rho_pn_phys_state}
\end{equation}
As can be easily seen from its form in Eq.~\eqref{eq:rho_pn_phys_state}, the superselected state $\rho^\text{P, N}(\theta,\varphi)$\ is separable, since it can be written as a simple mixture of product states. Furthermore, it is also classical since it is diagonal in a product basis. From this it follows, that all correlation in \eqref{eq:rho_pn_phys_state} between the two orbitals are classical. We summarize all correlation quantities with and without SSRs in Table \ref{tab:single}.

\begin{table}[h]
\begin{tabular}{|r|c|c|c|c|}
\hline
\rule{0pt}{2.6ex}\rule[-1.2ex]{0pt}{0pt}
 & $I$ & $C$ & $Q$ & $E$
\\
\hline
\rule{0pt}{2.6ex}\rule[-1.2ex]{0pt}{0pt}
No SSR & $2P(\theta)$ & $P(\theta)$ & $P(\theta)$ & $P(\theta)$
\\
\hline
\rule{0pt}{2.6ex}\rule[-1.2ex]{0pt}{0pt}
P/N-SSR & $P(\theta)$ & $P(\theta)$ & $0$ & $0$
\\
\hline
\end{tabular}
\caption{Total correlation $I$, classical correlation $C$, quantum correlation $Q$, and entanglement $E$ between the two orbitals $A$ and $B$ in the single electron state \eqref{eqn:single}, for the case without SSR and with P/N-SSR.}
\label{tab:single}
\end{table}

\subsection{Single Covalent Bond}\label{subsec:Hbond}

In this example, we apply the same quantum information concepts as in the previous section now to a pair of bonding electrons in a hydrogen-like diatomic molecule, described by the state
\begin{equation}
    |\Psi\rangle = f^\dagger_{\phi\uparrow}f^\dagger_{\phi\downarrow}|\Omega\rangle\ . \label{eqn:Hbond1}
\end{equation}
Here, $\phi$ is the bonding orbital, formed by superimposing two $1s$-like orbitals on the two nuclear centers $L$ (left) and $R$ (right)
\begin{eqnarray}
\phi = \mathcal{N} (\varphi_L + \varphi_R) \label{eqn:bonding_wf}
\end{eqnarray}
where $\mathcal{N}$ is a normalizing constant.

\begin{figure}[t]
    \centering
    \includegraphics[scale=0.5]{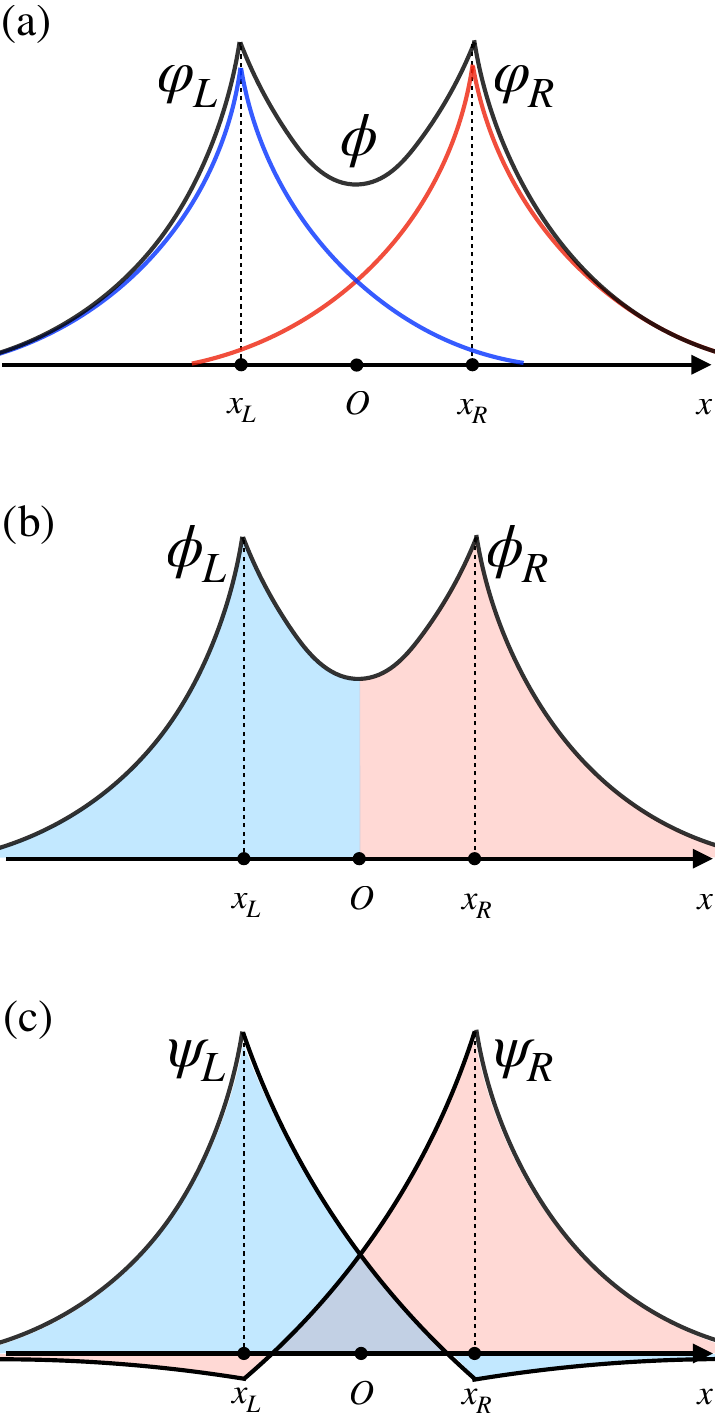}
    \caption{(a) Formation of the (up to normalization) bonding orbital $\phi$ with two local atomic $1s$-type orbitals $\varphi_L$ and $\varphi_R$ with nuclear centers at $x_{L,R}$. (b) Projection of the bonding orbital onto the left and right half space. (c) Rotated bonding and anti-bonding orbitals $\psi_{L}$ and $\psi_R$.}
    \label{fig:Hbond}
\end{figure}

Before we are able to proceed with a calculation of the correlation in $|\Psi\rangle$, being it quantum or classical, we first have to decide on a choice of orbital splitting. An obvious choice would be to consider the correlation between the bonding orbital $\phi$ and the corresponding anti-bonding orbital $\overline{\phi} = \overline{\mathcal{N}}(\varphi_L-\varphi_R)$. Together $\phi$ and $\bar\phi$ form a minimal active space, within which we will perform all our entanglement analysis. Referring to the splitting between $\phi$ and $\overline{\phi}$, $|\Psi\rangle$ is clearly a product state. As a result, it has zero correlation and entanglement. While this finding seems odd at first sight, as one would expect a considerable amount of entanglement to be ``stored" in a chemical bond, let us next consider a different, seemingly less intuitive alternate choice of the splitting.

In the top panel of Figure \ref{fig:Hbond}, we illustrate the formation of the bonding orbital $\phi$\ from the two local $1s$-type orbitals $\varphi_L$ and $\varphi_R$, respectively. We then make a cut at the center of the molecule dividing the space into left and right half, and project the bonding orbital onto the two half-spaces (central panel), denoted as $\phi_L$ and $\phi_R$. After normalization, the resulting two-electron wave function \eqref{eqn:Hbond1} can be written as
\begin{equation} \label{eqn:Hbond2}
    \begin{split}
        |\Psi\rangle &= \frac{1}{2} (f^\dagger_{\phi_L\uparrow} + f^\dagger_{\phi_R\uparrow})(f^\dagger_{\phi_L\downarrow} + f^\dagger_{\phi_R\downarrow}) |\Omega\rangle
        \\
        &= \frac{1}{2}(f^\dagger_{\phi_L\uparrow}f^\dagger_{\phi_L\downarrow}\! +\! f^\dagger_{\phi_R\uparrow}f^\dagger_{\phi_R\downarrow}\! +\! f^\dagger_{\phi_L\uparrow}f^\dagger_{\phi_R\downarrow} \!-\! f^\dagger_{\phi_L\downarrow}f^\dagger_{\phi_R\uparrow})|\Omega\rangle.
    \end{split}
\end{equation}
Hence, with respect to a splitting between the left and right projected orbitals $\phi_L$ and $\phi_R$, simple calculations lead to a set of entirely different values of correlation quantities of $\rho = |\Psi\rangle\langle\Psi|$
\begin{equation}
    \begin{split}
        \frac{1}{2} I(\rho) = Q(\rho) = C(\rho) = E(\rho) = 2\ln 2.
    \end{split}
\end{equation}

When superselection rules are considered, P-SSR and N-SSR eliminate coherent terms between different local parity and particle number sectors, respectively, leading to superselected states of the form
\begin{equation}
\begin{split}
        \rho^\text{P} &= \frac{1}{2} |\Psi_\text{even}\rangle\langle \Psi_\text{even}| + \frac{1}{2} |\Psi_\text{odd}\rangle\langle \Psi_\text{odd}|
        \\
        \rho^\text{N} &= \frac{1}{4}|\Psi_{2,0}\rangle\langle\Psi_{2,0}| + \frac{1}{4}|\Psi_{0,2}\rangle\langle\Psi_{0,2}|
        \\
         & \quad + \frac{1}{2}|\Psi_{1,1}\rangle\langle\Psi_{1,1}|,
\end{split}
\end{equation}
where
\begin{equation}
    \begin{split}
        |\Psi_\text{odd}\rangle &= |\Psi_{1,1}\rangle = \frac{1}{\sqrt{2}}(f^\dagger_{\phi_L\uparrow}f^\dagger_{\phi_R\downarrow} - f^\dagger_{\phi_L\downarrow}f^\dagger_{\phi_R\uparrow})|\Omega\rangle,
        \\
        |\Psi_\text{even}\rangle &= \frac{1}{\sqrt{2}}(f^\dagger_{\phi_L\uparrow}f^\dagger_{\phi_L\downarrow} + f^\dagger_{\phi_R\uparrow}f^\dagger_{\phi_R\downarrow}) |\Omega\rangle
        \\
        |\Psi_{2,0}\rangle &= f^\dagger_{\phi_L\uparrow}f^\dagger_{\phi_L\downarrow}|\Omega\rangle,
        \\
        |\Psi_{0,2}\rangle &=f^\dagger_{\phi_R\uparrow}f^\dagger_{\phi_R\downarrow}|\Omega\rangle.
    \end{split}
\end{equation}
In this case both P- and N-SSR reduce quantum correlation as well as entanglement by $50\%$ and $75\%$, respectively, while at the same time having no effect on the classical correlation. Consequently, the total correlation is lowered by the same amount of decrease in the quantum correlation, as the relation $I = C + Q$ holds in this case. We summarize the correlation quantities with and without SSRs in Table \ref{tab:Hbond}.

\begin{table}[h]
\begin{tabular}{|r|c|c|c|c|}
\hline
\rule{0pt}{2.6ex}\rule[-1.2ex]{0pt}{0pt}
$\phi, \overline{\phi}$ & $I$ & $C$ & $Q$ & $E$
\\
\hline
\rule{0pt}{2.6ex}\rule[-1.2ex]{0pt}{0pt}
No SSR & $0$ & $0$ & $0$ & $0$
\\
\hline
\rule{0pt}{2.6ex}\rule[-1.2ex]{0pt}{0pt}
P, N-SSR & $0$ & $0$ & $0$ & $0$
\\
\hline
\hline
\rule{0pt}{2.6ex}\rule[-1.2ex]{0pt}{0pt}
$\phi_L, \phi_R$ & $I$ & $C$ & $Q$ & $E$
\\
\hline
\rule{0pt}{2.6ex}\rule[-1.2ex]{0pt}{0pt}
No SSR & $4\ln 2$ & $2\ln 2$ & $2\ln 2$ & $2\ln 2$
\\
\hline
\rule{0pt}{2.6ex}\rule[-1.2ex]{0pt}{0pt}
P-SSR & $3\ln 2$ & $2\ln 2$ & $\ln 2$ & $\ln 2$
\\
\hline
\rule{0pt}{2.6ex}\rule[-1.2ex]{0pt}{0pt}
N-SSR & $\frac{5}{2}\ln 2$ & $2\ln 2$ & $\frac{1}{2}\ln 2$ & $\frac{1}{2}\ln 2$
\\
\hline
\end{tabular}
\caption{Total correlation, classical correlation, quantum correlation, and entanglement between the bonding and anti-bonding orbitals in single bond state $|\Psi\rangle$ in \eqref{eqn:Hbond1} (top panel), and between the two projected orbitals in the same state $|\Psi\rangle$ re-expressed in \eqref{eqn:Hbond2} (bottom panel), for the case without SSR, with P- and N-SSR.}
\label{tab:Hbond}
\end{table}

This example already illustrates that, by referring to a suitable orbital splitting that allows to capture a certain degree of spatial locality, one can recover strong correlation as one would expect in a chemical bond. Although the above choice of splitting seems from a quantum information perspective to be reasonable in terms of recovering correlation effects, it requires an artificial cut of the bonding orbital into two halves. Moreover, from the resulting orbitals $\phi_L$ and $\phi_R$, one cannot recover the anti-bonding orbital $\overline{\phi} \propto \varphi_L - \varphi_R$. As a matter of fact, $\{\phi_L, \phi_R\}$ do not span the same Hilbert space as the two local atomic orbitals $\{\varphi_L,\varphi_R\}$.
Therefore the optimal approach must include in addition the anti-bonding orbital into the total Hilbert space $\overline{\phi} = \overline{N} (\varphi_L - \varphi_R)$. To explore all possible choices of orbital bases, we unitarily (assuming for simplicity but without loss of generality real coefficients) transform the orbitals $\phi$, $\overline{\phi}$\
\begin{equation}
    \begin{split}
        \psi_L &= \cos(\theta) \phi + \sin(\theta) \overline{\phi},
        \\
        \psi_R &= -\sin(\theta) \phi + \cos(\theta) \overline{\phi}.
    \end{split}
\end{equation}
After this unitary basis rotation, we can rewrite the state $|\Psi\rangle$\ in Eq.~\eqref{eqn:Hbond1} as
\begin{equation} \label{eqn:Hbond3}
\begin{split}
    |\Psi\rangle &= [\cos^2(\theta) f^\dagger_{\psi_L\!\uparrow}f^\dagger_{\psi_L\!\downarrow} + \sin^2(\theta)f^\dagger_{\psi_R\!\uparrow}f^\dagger_{\psi_R\!\downarrow}
    \\
    &\quad + \cos(\theta)\sin(\theta)(f^\dagger_{\psi_L\!\uparrow}f^\dagger_{\psi_R\!\downarrow} -f^\dagger_{\psi_L\!\downarrow}f^\dagger_{\psi_R\!\uparrow})]|\Omega\rangle.
\end{split}
\end{equation}
The entanglement of $\rho = |\Psi\rangle\langle\Psi|$ is simply
\begin{equation}
\begin{split}
    E(\rho) &= -2[\cos^2(\theta)\ln(\cos^2(\theta))\!+\!\sin^2(\theta)\ln(\sin^2(\theta))]
    \\
    &= 2P(\theta).
\end{split}
\label{eqn:Hbond3-E}
\end{equation}
From Eq.~\eqref{eqn:Hbond3-E} it follows that maximal entanglement is realized by a rotation with angle $\theta = \frac{\pi}{4}$. As can be seen from Table \ref{tab:Hbond}, in the latter basis the resulting $E(\rho)$\ reaches also $2\ln 2$, in perfect agreement with the case of the artificial half-splitting discussed previously. Moreover, assuming a rotation of $\theta=\frac{\pi}{4}$, the transformed orbitals $\psi_{L}$ and $\psi_R$, illustrated in the bottom panel (c) of Figure \ref{fig:Hbond}, are simply equal superpositions of the initial bonding $\phi$\ and anti-bonding $\overline{\phi}$\ orbitals, with plus and minus signs respectively. As a matter of fact, the final expression for the state in Eq.~\eqref{eqn:Hbond3}, expressed in the basis $\{\psi_{L},\psi_R\}$, takes the same form as its counterpart in Eq.~\eqref{eqn:Hbond2}, with the replacement $\psi_{L,R} \rightarrow \phi_{L,R}$. Therefore, in the particular choice of $\theta = \frac{\pi}{4}$, we find that all correlation quantities (with or without SSRs) of the state given by Eq.~\eqref{eqn:Hbond3}\ coincide with those of the state in Eq.~\eqref{eqn:Hbond2}. Thus, we can interpret the rotation angle $\theta=\frac{\pi}{4}$ as the point where maximal orbital localization effect is achieved, while still keeping the orbitals orthogonal and without dissecting them.

To explore more comprehensively the connection between orbital entanglement and chemical bonding, let us consider the cases of maximal and minimal entanglement in some prototypical states of definite bond order. In molecular orbital (MO) theory, the bond order of a state is defined as the difference in the occupation number between the bonding orbital $\phi$ and its anti-bonding orbital partner $\overline{\phi}$ divided by $2$\cite{mqm_atkins}
\begin{equation}
    \text{bond order} = \frac{1}{2}(N_{\text{bond}}-N_{\text{antibond}}). \label{eqn:BO}
\end{equation}
To illustrate the concept of a bond order we consider the following four states
\begin{equation}\label{eqn:definite_bonds}
\begin{split}
&|\Psi_1\rangle = f^\dagger_{\phi \uparrow} |\Omega\rangle,
\\
&|\Psi_2\rangle = f^\dagger_{\phi \uparrow} f^\dagger_{\phi \downarrow} |\Omega\rangle,
\\
&|\Psi_3\rangle = f^\dagger_{\phi \uparrow} f^\dagger_{\phi \uparrow} f^\dagger_{\overline{\phi} \uparrow} |\Omega\rangle,
\\
&|\Psi_4\rangle = f^\dagger_{\phi \uparrow}f^\dagger_{\phi \downarrow}  f^\dagger_{\overline{\phi} \uparrow} f^\dagger_{\overline{\phi} \downarrow} |\Omega\rangle,
\end{split}
\end{equation}
which have a bond order of $\frac{1}{2}$, $1$, $\frac{1}{2}$, and $0$, respectively, according to \eqref{eqn:BO}. We can easily find the minimal entanglement of all four states to be zero, with respect to the orbital partition between $\phi$ and $\overline{\phi}$, that is, the bonding and anti-bonding orbitals. Under an arbitrary orbital rotation of angle $\theta$
\begin{equation}
    \begin{split}
        f^{\dagger}_{\psi_L\sigma} &= \cos(\theta) f^\dagger_{\phi\sigma} + \sin(\theta) f^\dagger_{\overline\phi\sigma},
        \\
        f^{\dagger}_{\psi_R\sigma} &= \sin(\theta) f^\dagger_{\phi\sigma} - \cos(\theta) f^\dagger_{\overline\phi\sigma},
    \end{split}
\end{equation}
the four states transform to
\begin{equation} \label{eqn:bond_order}
\begin{split}
    &|\Psi_1\rangle \!=\! \cos(\theta)f^\dagger_{\psi_L\!\uparrow}|\Omega\rangle + \sin(\theta)f^\dagger_{\psi_R\!\uparrow}|\Omega\rangle,
    \\
    &|\Psi_2\rangle \!=\! \cos^2(\theta)f^\dagger_{\psi_L\!\uparrow}f^\dagger_{\psi_L\!\downarrow}|\Omega\rangle + \sin^2(\theta)f^\dagger_{\psi_R\!\downarrow}f^\dagger_{\psi_R\!\downarrow}|\Omega\rangle
    \\
    \nonumber &\quad \quad \quad \cos(\theta)\sin(\theta)(f^\dagger_{\psi_L\!\uparrow}f^\dagger_{\psi_R\!\downarrow}-f^\dagger_{\psi_L\!\downarrow}f^\dagger_{\psi_R\!\uparrow})|\Omega\rangle,
    \\
    &|\Psi_3\rangle \!=\!f^\dagger_{\psi_L\!\uparrow}( \cos(\theta)  f^\dagger_{\psi_L\!\downarrow} f^\dagger_{\psi_R\!\uparrow} \!-\! \sin(\theta) f^\dagger_{\psi_R\!\uparrow} f^\dagger_{\psi_R\!\downarrow})|\Omega\rangle,
    \\
    &|\Psi_4\rangle \!=\! f^\dagger_{\psi_L\!\uparrow} f^\dagger_{\psi_L\!\downarrow} f^\dagger_{\psi_R\!\uparrow} f^\dagger_{\psi_R\!\downarrow}|\Omega\rangle.
\end{split}
\end{equation}
The latter shows that the resulting entanglement $E$ with respect to the partition between $\psi_L$ and $\psi_R$ of $|\Psi_{1,2,3}\rangle$ is maximized for $\theta=\frac{\pi}{4}$, whereas the entanglement of $|\Psi_4\rangle$ remains invariant under orbital transformation. To summarize these findings we compile in Table \ref{tab:bond_order} the maximal and minimal entanglement of these four states as a function of the bond order. Remarkably, we find that the maximal entanglement, realized between the maximally localized orbitals, is \textit{indeed} proportional to the bond order \eqref{eqn:BO} of each state. A single bond of bond order $1$ thus corresponds to the entanglement value $E = 2\ln 2$ between the fully-localized atomic-like orbitals. Intriguingly, this value exceeds by an order of magnitude the numbers reported in previous studies  \cite{Reiher13,Reiher14,Reiher15,Ayers15a,Ayers15b,Bogus15,Szalay17,Reiher17b,Legeza18,Legeza19}. In turn, this clearly demonstrates that quantum information tools applied to delocalized orbitals describe primarily the validity of the independent electron-pair picture rather than the bonding structure of molecular systems.

\begin{table}[t]
    \centering
    \begin{tabular}{|r|c|c|c|c|}
    \hline
    \rule{0pt}{2.6ex}\rule[-1.2ex]{0pt}{0pt}
        $\overline\phi$ & $\msout{\phantom{\uparrow\downarrow}}$ & $\msout{\phantom{\uparrow\downarrow}}$ & $\msout{\uparrow\!\!\phantom{\downarrow}}$ & $\msout{\uparrow\downarrow}$
        \\
        \rule{0pt}{2.6ex}\rule[-1.2ex]{0pt}{0pt}
        ${\phi}$ & $\msout{\uparrow\!\!\phantom{\downarrow}}$ & $\msout{\uparrow\downarrow}$ & $\msout{\uparrow\downarrow}$ & $\msout{\uparrow\downarrow}$
        \\
    \rule{0pt}{2.6ex}\rule[-1.2ex]{0pt}{0pt}
    & $|\Psi_1\rangle$ & $|\Psi_2\rangle$ & $|\Psi_3\rangle$ & $|\Psi_4\rangle$
    \\
        \hline
        \rule{0pt}{2.6ex}\rule[-1.2ex]{0pt}{0pt}
        Bond Order & $\frac{1}{2}$ & $1$ & $\frac{1}{2}$ & $0$
        \\
        \hline
        \rule{0pt}{2.6ex}\rule[-1.2ex]{0pt}{0pt}
        $E_\text{max}$ & $\ln 2$ & $2\ln 2$ & $\ln 2$ & $0$
        \\
        \hline
        \rule{0pt}{2.6ex}\rule[-1.2ex]{0pt}{0pt}
        $E_\text{min}$ & $\phantom{\ln2}0\phantom{\ln2}$ & $\phantom{\ln2}0\phantom{\ln2}$ & $\phantom{\ln2}0\phantom{\ln2}$ & $\phantom{\ln2}0\phantom{\ln2}$
        \\
        \hline
    \end{tabular}
    \caption{Bond order, maximal and minimal entanglement of the four states given in Eq.~\eqref{eqn:definite_bonds}.}
    \label{tab:bond_order}
\end{table}

Lastly, we remark that the insights we gained from the single covalent bond states extends beyond bonds of order 1. Namely, for a prototypical $K$-fold bond state $|\Psi_K\rangle = \prod_{k=1}^{K} f^\dagger_{\phi_k\uparrow}f^\dagger_{\phi_k\downarrow}|0\rangle$, rotating pairs of bonding and antibonding orbitals $\phi_k$ and $\bar{\phi}_k$ by $\pi/4$ leads to $K$ pairs of maximally entangled rotated orbitals, amounting to $K(2\ln2)$ of total orbital-orbital entanglement.

\section{Computational details}\label{sec:comput}

In Section \ref{sec:results}, we will analyze and decompose electron-correlation effects into its classical and quantum correlation as well as entanglement contributions at the example of a few chain-like and cyclic $\pi$-conjugated organic molecules in their electronic ground state, namely, ethylene (C$_2$H$_4$), decapentaene (C$_{10}$H$_{12}$), eicosadecaene (C$_{20}$H$_{22}$), and benzene (C$_6$H$_6$). In this preceding section, we will present the computational details of our ground state calculation, as well as an algorithm for numerically obtaining the quantum correlation.

\subsection{General Considerations}\label{subsec:comput:general}

For each molecular compound, the geometrical parameters were taken from literature and are listed as xyz coordinates in the Appendix (see Tables \ref{tab:benzene:coord}-\ref{tab:eicosadecaene:coord}). Since our primary focus will be on rationalizing the chemical bonding as well as electron correlation effects that originate from the $\pi$-subspace of the above mentioned conjugated systems,
we carried out complete-active-space (CAS) calculations correlating $n_e$\ electrons in $n_o$\ $\pi$-orbitals.
In this particular case it follows that $n_e=n_o$\ which in turn equals the number of carbon atoms in the molecule.
Hence, by choice, our analysis neglects in the present work any electron correlation contributions arising from $\sigma$-type MOs as well as set(s) of higher-lying (correlating) secondary $\pi^*$-orbitals.
To corroborate this approximation, we performed additional test calculations on decapentaene considering CAS spaces of CAS(34,34) and CAS(10,20), respectively.
Those large-scale test calculations  revealed that correlation contributions to the valence $\pi$-$\pi^\ast$-space arising either from the C-C and C-H $\sigma$-space (CAS(34,34)) or from additional correlating $\pi^*$-orbitals (CAS(10,20)) are (i) differential (within the additional sets of $\{\sigma,\sigma^\ast\}$ MOs) or (ii) safely negligible with natural orbital occupation numbers of the valence $\pi$-$\pi^\ast$-space changing by less than $\pm$0.003.

All quantum-chemical calculations, except for the density matrix renormalization group (DMRG) calculations\cite{whit92,whit93,schol11} (for a recent review of DMRG in quantum chemistry, see for example Ref.~\onlinecite{baia20a}) further described below,
were performed with the 2019 version of the \textsc{Molpro} software package\cite{molpro1,molpro2,molpro3}. By making use of the \texttt{FCIDUMP} file format\cite{know89b}\ in \textsc{Molpro}, we exported the (effective) one- and two-electron Hamiltonian integrals in a given MO basis for the ensuing MPS wave function optimization within the DMRG software \textsc{QCMaquis}\cite{kell15a,kell16,knec16a}. All calculations were carried out in C$_1$ point group symmetry as well as with correlation-consistent Dunning-type basis sets\cite{dunn89} of double-$\zeta$ quality (\texttt{cc-pVDZ}).
The latter one-particle basis set should be sufficiently large to provide a qualitatively correct description of the valence correlation effects, in particular of the $\pi$-bonds.
In order to critically assess the correlation contributions introduced in Section \ref{subsec:Corrtypes} within the $\pi$-space manifold of our molecular systems, we considered three distinct set of MOs that are related to each other by suitable orbital rotations.
To this end, we first performed self-consistent field (SCF) Hartree-Fock (HF) calculations for the spin-singlet ($S=0$) ground state of each molecule, yielding a set of \emph{canonical} HF MOs for the respective molecular systems. In an ensuing step, we then applied a Pipek-Mezey (PM)\cite{pipe89} localization procedure which yields a second set of \emph{localized}\ MOs (dubbed as ``PM-localized'' in the following) while retaining the $\sigma$-\ and $\pi$-character of the initial canonical MOs, respectively. It should be emphasized that, although it is possible to perform the PM-localization on the entire set of canonical MOs\cite{Szalay17}, in this work the PM-localization is implemented separately within the bonding (occupied) and antibonding (virtual) orbital sub-manifolds, as it is common practice in the quantum chemistry community. The final set of \emph{atomic-like} of orbitals was obtained by means of a (sequence of) $2\times2$ Jacobi-rotation(s) by an angle $\theta$ within the $\pi$-space manifold and setting out either from the PM-localized MO basis (ethylene and polyenes) or the canonical MO basis (benzene).
In the former, a single rotation by $\theta=\frac{\pi}{4}$ within a pair of bonding $\pi$ and its antibonding partner $\pi^\ast$ suffices to yield the desired atomic-like molecular basis. It is worthwhile to mention that the latter scheme was recently also explored in an attempt to reduce the 1-norm of the Hamiltonian in the context of quantum computing applications\cite{kori21a}.
In contrast to the remaining molecules studied in this work, obtaining an atomic-like orbital basis for benzene requires in general six-orbital unitary transformations, which can be decomposed into three consecutive sets of pairwise rotations (see Appendix \ref{app:benzene}).

In order to (approximately) solve the full CI problem for a given CAS orbital space, we employed a spin-adapted DMRG algorithm\cite{kell16}\ as implemented in \textsc{QCMaquis}\cite{kell15a,kell16,knec16a}.
Having thus obtained an optimized MPS wave function for the singlet electronic ground state, the latter was then used to compute the one- and two-\emph{orbital} reduced density matrices which are the primary input quantities for the correlation measures defined in Section \ref{subsec:Corrtypes}. In all DMRG calculations we employed a two-site optimization algorithm starting from a HF guess (\texttt{init\_guess=hf} in \textsc{QCMaquis}) for the initial MPS while varying the maximum number of renormalized block states $m$ from $m=500$ up to $m=2000$ until the total energy was converged to at least less than sub-$\mu$ Hartree accuracy. Furthermore, all DMRG calculations were carried out with an orbital ordering of pairwise correlating $\pi$-$\pi^\ast$-orbitals.

For the orbital visualizations we made use of \textsc{Jmol}\cite{jmol}. To aid the reader's visual guide, we applied for each molecular system individual isosurface thresholds as indicated in the figures.

\subsection{Numerical calculation of quantum correlation}\label{subsec:algClCor}

The set of classically correlated states \eqref{eqn:Dcl} has a complicated and highly non-convex structure, which makes an optimization over it a formidable task. Fortunately, Ref.~\cite{modi2010unified} provides a suitable theorem that connects the spectrum of the closest classical state $\chi_\rho$ to the diagonal entries of $\rho$ in the eigenbasis of $\chi_\rho$. More precisely, if $\chi_\rho = \sum_{ij}\lambda_{ij}|i\rangle\langle i|\!\otimes\! |j\rangle\langle j|$ is the closest classical state to $\rho$, then its spectrum is given by
\begin{equation}
    \lambda_{ij} = \langle i|\!\otimes\! \langle j| \,\rho\, |i\rangle \!\otimes\! |j\rangle.
\end{equation}
In other words, the closest classical state to $\rho$ is of the form
\begin{equation}
    \chi_\rho=\sum_{ij}|i\rangle\langle i|\!\otimes\! |j\rangle\langle j| \rho |i\rangle\langle i|\!\otimes\! |j\rangle\langle j|. \label{eqn:CCS}
\end{equation}
This finding represents the starting point for our quest to search for the optimal local bases $\{|i\rangle\}$ and $\{|j\rangle\}$ of two subsystems $A$ and $B$, respectively, recovering the minimizer of \eqref{eqn:Q}. Given that any two bases can be connected by a unique unitary operator $U|i\rangle \mapsto |i'\rangle$, provided that we fix the local computational bases, this search is then equivalent to finding the optimal unitary operators $U_A$ and $U_B$ for the respective subsystems. In the following we present a random walk algorithm assisted by probabilistic rejection, in search for the optimal local unitary operators $U_A$ and $U_B$ within the manifolds of local unitaries $\mathcal{U}_A$ and $\mathcal{U}_B$, respectively.

\begin{algorithm}[t] \label{alg:disc}
\caption{Calculating Quantum Correlation}
\SetAlgoLined

\textbf{INPUT:} bipartite quantum state $\rho$

\textbf{OUTPUT:} $Q(\rho)$ and the closest classical state $\chi_\rho$ to $\rho$

\textbf{COMPUTATION:}

\textbf{SET} initial local bases $\{|i^{(0)}\rangle_A\}$ and $\{|j^{(0)}\rangle_B\}$;

\textbf{SET} $n = 0$ and $U^{(0)}_A = U^{(0)}_B = \openone$;

\textbf{COMPUTE} $\chi^{(0)}_\rho = \sum_{ij}|i^{(0)}\rangle\langle i^{(0)}|\!\otimes\! |j^{(0)}\rangle\langle j^{(0)}| \rho |i^{(0)}\rangle\langle i^{(0)}|\!\otimes\! |j^{(0)}\rangle\langle j^{(0)}|$;

\textbf{COMPUTE} $Q(\rho)^{(0)} = S(\rho||\chi^{(0)}_\rho)$;

\textbf{WHILE} $n < N_\text{step}$ \textbf{DO}:

\quad\quad \textbf{SAMPLE} random unitary matrices $V_{A,B}$;

\quad\quad\textbf{COMPUTE} $V_{A,B} \leftarrow V_{A,B}^{\frac{1}{M}}$;

\quad\quad\textbf{UPDATE} $U_{A,B}^{(n+1)} \leftarrow V_{A,B} U^{(n)}_{A,B}$;

\quad\quad\textbf{COMPUTE} new local bases

\quad \quad \quad $U_A^{(n+1)} |i^{(n)}\rangle \mapsto |i^{(n+1)}\rangle$, $U_B^{(n+1)} |j^{(n)}\rangle \mapsto |j^{(n+1)}\rangle$

\quad\quad\textbf{COMPUTE} new classical state

\quad \quad \quad $\chi^{(n+1)}_\rho = \sum_{ij}|i\rangle\langle i|\!\otimes\! |j\rangle\langle j| \rho |i\rangle\langle i|\!\otimes\! |j\rangle\langle j|$;

\quad\quad\textbf{COMPUTE} $Q^{(n+1)} = S(\rho||\chi^{(n+1)})$;

\quad\quad\textbf{SAMPLE} uniformly $p \in (0,1]$;

\quad\quad\textbf{IF} $Q^{(n+1)}<Q^{(n)}$:

\quad\quad\quad\quad \textbf{UPDATE} $Q(\rho) \leftarrow Q^{(n)}(\rho)$;

\quad\quad\quad\quad \textbf{UPDATE} $\chi_\rho \leftarrow \chi^{(n)}_\rho$;

\quad\quad\quad\quad \textbf{UPDATE} $n \leftarrow n + 1$;

\textbf{END}
\end{algorithm}

The computational scheme outlined in Algorithm~\ref{alg:disc}\ consists of the following steps: One first initializes a pair of local bases sets $\{|i^{(0)}\rangle\}$ and $\{|j^{(0)}\rangle\}$ for the two subsystems as well as the two local unitary operators as $U^{(0)}_{A,B} = \openone$. The initial bases determine a candidate for the closest classical state $\chi^{(0)}_\rho$ according to \eqref{eqn:CCS} and the distance $Q^{(0)} = S(\rho||\chi^{(0)}_\rho)$. The unitary operators live on the connected manifolds $\mathcal{U}_{A,B} \ni U_{A,B}$ which allows us to make use of a random walk algorithm to find the optimal set of unitary operators.
We start by performing a small step in $\mathcal{U}_{A,B}$ by multiplying $U^{(0)}_{A,B}$ with a ``small'' unitary operator $V_{A,B}$ close to the identity, arriving at $U^{(1)}_{A,B}$, which, in turn, determines a pair of local bases $U^{(1)}_{A}|i^{(0)}\rangle \equiv |i^{(1)}\rangle$, $U^{(1)}_{B}|j^{(0)}\rangle \equiv |j^{(1)}\rangle$. If these new bases define a closer classical state according to \eqref{eqn:CCS}, then this step is accepted, and otherwise rejected. The latter enables us to avoid being trapped by a local minimum. This procedure is repeated until (i) a desired accuracy or (ii) a predefined number of steps is reached. To compensate for the stochastic nature of Algorithm~\ref{alg:disc}, 10 initial local bases sets are chosen and the closest resulting classical state is taken as the optimal one. The step size parameter $M$ is chosen to be $10^3$, and the number of steps $N_\text{step} = 10^{4}$.

\section{Numerical results}\label{sec:results}

In this section, we present the main results of this paper, namely the correlation and entanglement pattern in the $\pi$-bonds of the molecular electronic ground states of $\pi$-conjugated organic molecules.

\subsection{Ethylene} \label{subsec:ethylene}

\begin{figure}[t]
    \centering
    \includegraphics[scale=0.6]{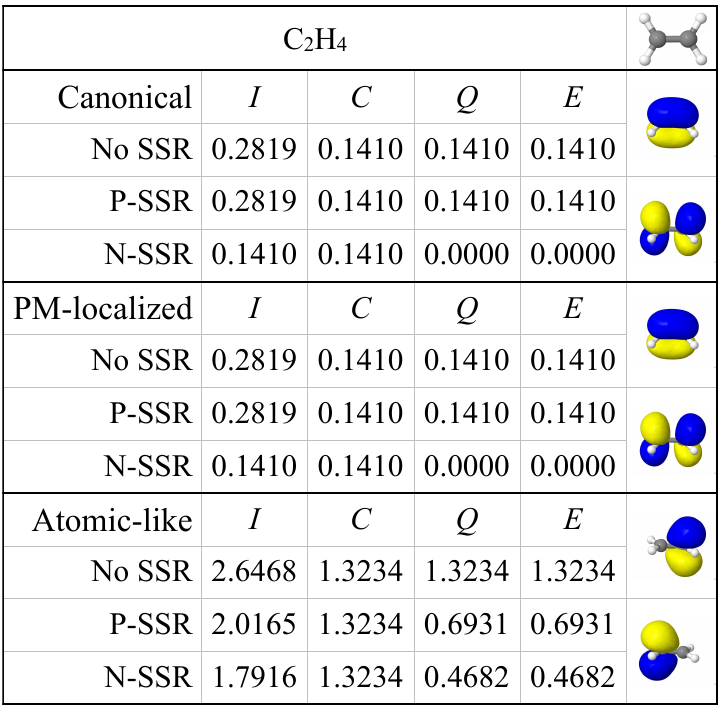}
    \caption{Total correlation $I$, classical correlation $C$, quantum correlation $Q$, and entanglement $E$ between two canonical, PM-localized, and atomic-like orbitals used for constructing the ground state of C$_{2}$H${_4}$, for the case without SSR, with P-SSR, and with N-SSR. For each choice of orbital pair, contour plots (with an isosurface value of 0.05) of the orbitals are shown on the right most column.}
    \label{fig:C2H4_s2}
\end{figure}

We first study one of the simplest molecules containing a prototypical $\pi$-bond, namely ethylene ($\mathrm{C_2H_4}$). Since the $\pi$-bond comprises only two carbon centers, the resulting CAS contains only $n_e\!=\!n_o\!=\!2$ or for short CAS(2,2), that is the bonding and anti-bonding orbitals $\pi$ and $\pi^\ast$, which are constructive and destructive superposition of two $p_z$-orbitals (assuming bonding along the $z$-axis) on the two carbon atoms, respectively. For non-interacting electrons, both electrons would occupy the energetically more favorable $\pi$-orbital, forming a product state with respect to partitioning of the bonding and antibonding orbitals, as in \eqref{eqn:Hbond1}, whereas electron interaction lifts the occupancy of the $\pi^\ast$-orbital to around $0.03$ electron pair, thus introducing a small deviation from the aforementioned product state. From this simple observation, we should expect a low correlation and entanglement between the $\pi$- and $\pi^\ast$-orbital, and this is indeed what we conclude from our analysis in Figure \ref{fig:C2H4_s2}\ for the common, \textit{canonical} case. However, more correlation and entanglement will be recovered as we \textit{fully} localize these two orbitals.

In Figure \ref{fig:C2H4_s2} we listed the total correlation $I$, classical correlation $C$, quantum correlation $Q$, and entanglement $E$ between the two orbitals based on canonical, PM-localized, and atomic-like orbitals. In addition, for each correlation quantity, the effect of P-SSR and N-SSR are taken into account in our analysis. Since the active space comprises in all three cases only two orbitals, it follows that single orbital correlation and entanglement (those between one orbital and the rest of the system) will coincide with the orbital-orbital ones.

Starting with the canonical case (upper panel of Figure~\ref{fig:C2H4_s2}), all correlation quantities are low, as anticipated above. In the case without SSRs and recalling that in this particular example the two-orbital reduced state is pure, the total correlation $I$ is simply twice the amount of the classical $C$ and quantum parts $Q$ such that there is no distinction between quantum correlation $Q$ and entanglement $E$. Interestingly, P-SSR does not show any diminishing effect in any of the correlation quantities. We can explain this somewhat surprising fact with the observation that the electronic ground state does not contain contributions from singly-excited configurations, and thus can solely be written as a superposition of doubly-occupied and empty configurations, respectively. Moreover, this is also the reason why no quantum correlation or entanglement survive in the presence of N-SSR: the ground state contains only a superposition of configurations with different local particle numbers.

Hence, in order to recover strong correlation and entanglement in the $\pi$-bond, we applied two different localization schemes to the canonical orbitals as detailed in Section \ref{subsec:comput:general}. In this particular case, the PM-localization scheme yields \emph{localized} $\pi$-MOs matching the original canonical $\pi$-orbitals, since the system comprises only $\sigma$- and $\pi$-type orbitals and the PM-localization scheme preserves the $\sigma$- and $\pi$-character of the MOs. As a result, we see no difference in their respective correlation quantities with respect to the canonical case and the data for the PM case coincide in all three rows (see second panel in Figure~\ref{fig:C2H4_s2}) with their respective counterpart in the upper panel of the canonical case. In passing we note that these findings will not hold when the canonical MOs span over several bonding regions, as we shall see in the following sections.
Our second localization scheme sets out from the PM-localized MOs (in this case equivalently the canonical MOs) where we apply in an ensuing step a $2\times2$ Jacobi rotation between the two PM-localized orbitals by an angle $\theta=\frac{\pi}{4}$. This unitary rotation leads to two atomic-like orbitals
\begin{subequations}
\begin{align}
    \psi_L = \frac{1}{\sqrt{2}} \pi_\text{PM} + \frac{1}{\sqrt{2}} \pi^\ast_\text{PM},
    \\
    \psi_R = \frac{1}{\sqrt{2}} \pi_\text{PM} - \frac{1}{\sqrt{2}} \pi^\ast_\text{PM},
\end{align}
\end{subequations}
where $\pi^{(\ast)}_\text{PM}$ are the PM-localized $\pi^{(\ast)}$-orbitals. As illustrated in the third panel of Figure \ref{fig:C2H4_s2} (atomic-like), the resulting $\psi_{1,2}$ are indeed \emph{fully localized} around one carbon center in stark contrast to the localized molecular orbitals obtained from the PM localization scheme. Moreover, these atomic-like orthogonal orbitals act \textit{identically} as the original atomic orbitals, in a sense that the same linear combination of the former as the latter give rise to the bonding and antibonding orbitals (up to overall normalization)
\begin{equation}
    \begin{split}
        \pi_{PM} = \frac{1}{\sqrt{2}} \psi_L + \frac{1}{\sqrt{2}}\psi_R,
        \\
        \pi^\ast_{PM} = \frac{1}{\sqrt{2}} \psi_L - \frac{1}{\sqrt{2}}\psi_R,
    \end{split}
\end{equation}
thus preserving the information of the bond construction. And most importantly, we now recover strong correlation and entanglement in the $\pi$-bond. Without SSRs, the entanglement $E$\ between $\psi_1$ and $\psi_2$\ reaches $95\%$\ of its maximum value of $2\ln 2$, in excellent agreement with the degree of entanglement that we observed for a prototypical bond in the analytic example discussed in Section \ref{subsec:Hbond}. Moreover, the effect of SSRs is qualitatively different in case of the atomic-like orbitals. P-SSR cancels around half of the entanglement, whereas around a third of the entanglement is still accessible under the restriction of N-SSR. The latter is rooted in the complexity of the ground state wave function which, in contrast to the much simpler form within the canonical and PM-localized orbitals bases, is now composed of several configurations of comparable weights, including those with single local occupations.

It is worth noting that the fully localized orbitals are only $95\%$ maximally entangled. This deviation from the perfect single bond in Section III B is not an artefact of an imperfect choice of orbitals, but rather an inevitable consequence of electron interaction. The latter namely introduces a multireference character to the ground state, and excites finite occupation in the antibonding orbital. In other words, the maximal entanglement $2\ln2$ in a perfect single bond state $f^\dagger_{\phi\uparrow}f^\dagger_{\phi\downarrow}|0\rangle$ can never be realized in an interacting molecule.

To summarize the main conclusions from this seemingly simple example, enforcing \textit{atomic-like}\ locality in the MO basis for the $\pi$-orbital space leads to two distinct features of the $\pi$-bond in comparison to the commonly considered canonical case: (i) the ground state electronic wave function markedly changes character from single- to strongly multi-configurational and, more importantly, (ii) the actual entanglement $E$ between the valence $\pi^{(\ast)}$-orbitals without SSR increases drastically from 5\% to 95\% of its maximum value of $2\ln 2$ which was established by means of an analytical model for a chemical bond in Section \ref{subsec:Hbond}.

\subsection{Polyene}\label{subsec:polyene}

Having analyzed in the previous section the conceptually most simple ``mono"-ene, we will focus in the following on all-\textit{trans} polyenes CH$_2$-(CH)$_n$-CH$_2$, a family of extended, prototypical, $\pi$-conjugated molecular systems. More specifically, we consider two exemplary systems with $n=8$ (decapentaene, $\mathrm{C_{10}H_{12}}$) and $n=18$ (eicosadecaene, $\mathrm{C_{20}H_{22}}$). To unambiguously study the individual correlation contributions within the valence $\pi^{(\ast)}$-space requires for those molecular systems active orbital spaces of CAS(10,10) and CAS(20,20), respectively. Given the size of these CAS spaces, single-orbital and orbital-orbital correlations will no longer coincide, and need to be addressed separately. With $n_o > 2$, the single orbital correlation quantifies the correlation between one orbital and all other orbitals, including multipartite correlations, much beyond any orbital-orbital correlations. As in the case of ethylene in the previous section, we will consider for our analysis three distinct choices of MO basis, namely canonical, PM-localized, and atomic-like orbitals.

\begin{figure*}[htb]
    \centering
    \includegraphics[width=0.8\textwidth]{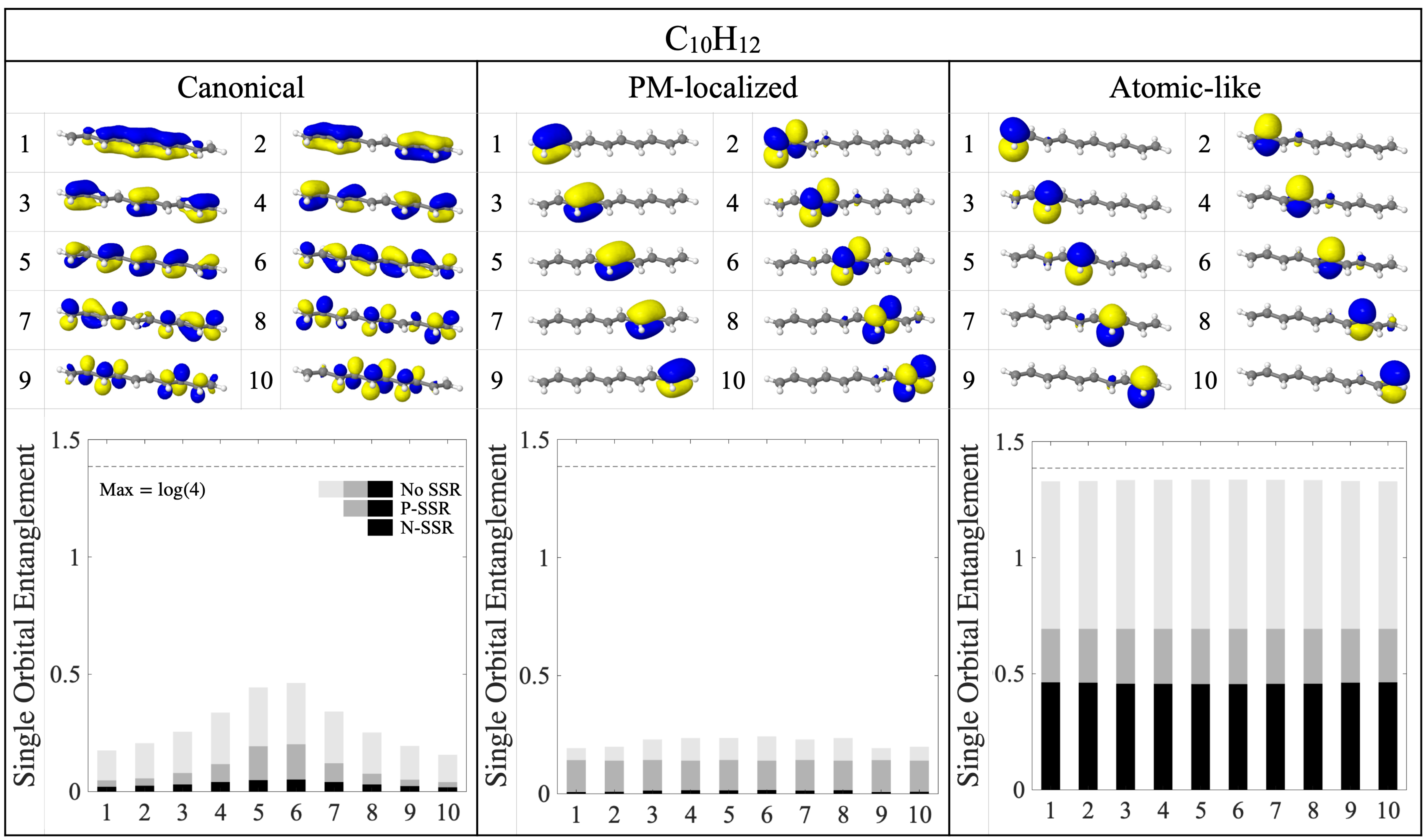}
    \caption{Single orbital entanglement in the CAS(10,10)-optimized electronic ground state of $\mathrm{C_{10}H_{12}}$. The orbital numbering on the $x$-axis in the lower panel follows the one given for the canonical, PM-localized, and atomic-like orbitals in the upper panel (plotted with an isosurface value of 0.05). The color code for the single orbital entanglement data is as follows: no SSR (all color), P-SSR (black and dark grey), and N-SSR (black).}
    \label{fig:C10H12_s1}
\end{figure*}

\begin{figure*}[htb]
    \centering
    \includegraphics[width=0.8\textwidth]{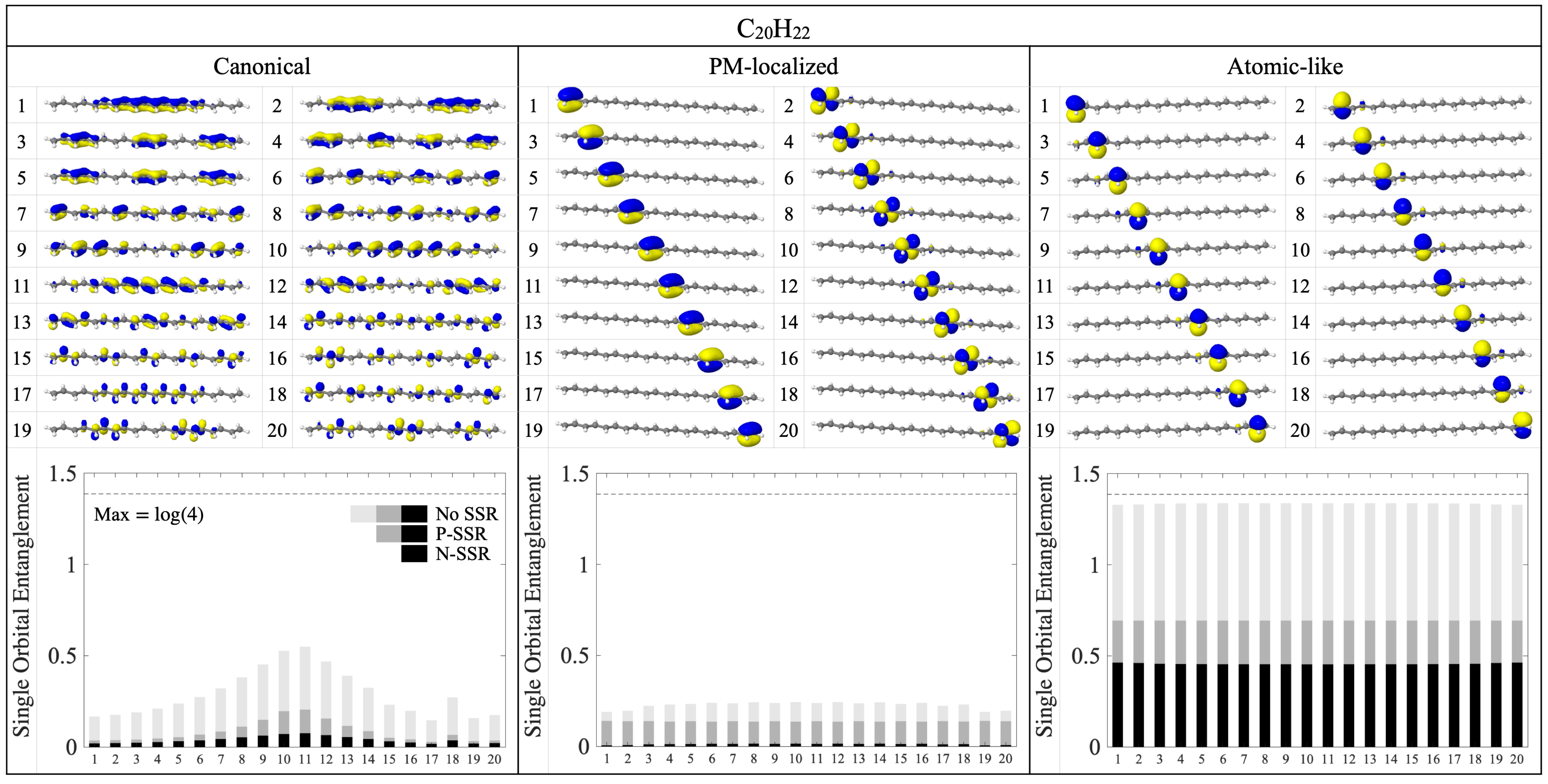}
    \caption{Single orbital entanglement in the CAS(20,20)-optimized electronic ground state of $\mathrm{C_{20}H_{22}}$. The orbital numbering on the $x$-axis in the lower panel follows the one given for the canonical, PM-localized, and atomic-like orbitals in the upper panel (plotted with an isosurface value of 0.05). The color code for the single orbital entanglement data is as follows: no SSR (all color), P-SSR (black and dark grey), and N-SSR (black).}
    \label{fig:C20H22_s1}
\end{figure*}

\subsubsection{Single Orbital Correlation} \label{subsubsec:single}

As we are dealing with a pure ground state, single orbital correlation can be related to the single orbital entanglement via simple linear relations\cite{ding2020concept}, and the latter is equivalent with the single-orbital quantum correlation. To this end, it suffices to focus in this paragraph solely on single-orbital entanglement data.

In Figure \ref{fig:C10H12_s1} and Figure \ref{fig:C20H22_s1}, respectively, we present the canonical, PM-localized, and atomic-like orbitals for $\mathrm{C_{10}H_{12}}$ and $\mathrm{C_{20}H_{22}}$ (upper panels) along with the single orbital entanglement of each orbital (lower panels of Figures \ref{fig:C10H12_s1} and \ref{fig:C20H22_s1}), namely the entanglement between one orbital and the remaining orbitals comprised in the active space.
Before embarking on an in-depth discussion of the entanglement data, we first emphasize two obvious key differences between the MOs of the polyenes shown in the upper panels of Figure \ref{fig:C10H12_s1} and Figure \ref{fig:C20H22_s1}, respectively, and those of ethylene: (i) the canonical MOs for both extended systems are highly delocalized across the entire carbon-carbon chain; (ii) the PM-localized MOs no longer coincide with the canonical ones, and are localized only around two carbon centers involved in a $\pi^{(\ast)}$-bond. Hence, the PM-localization scheme succeeds in \emph{partially} localizing the canonical MOs. Finally, as was the case for ethylene, to obtain atomic-like orbitals  requires a further rotation of each corresponding pair of PM-localized $\pi^{(\ast)}$-MOs by $\theta=\frac{\pi}{4}$.

Considering next the single orbital entanglement shown in the lower panels of the respective Figures \ref{fig:C10H12_s1} and \ref{fig:C20H22_s1}, the three choices of MO basis reveal drastically different behaviors. For both molecules, the canonical MOs display a large variation in their single orbital entanglement. The most entangled orbital \#6 (\#11 in $\mathrm{C_{20}H_{22}}$) corresponds to the LUMO in either case and contains almost three times the amount of the least entangled ones. In addition, we observe that both P-SSR and N-SSR drastically reduce the entanglement available in the canonical MOs. The fact that we still find single orbital entanglement for the $\pi$-orbitals is therefore a clear indication of their departure from a double occupancy (as would be expected in an uncorrelated mean-field model) in the CAS-optimized ground-state wave function due to the presence of electron-electron correlation which is most dominant for the HOMO and HOMO-1. Likewise, a similar explanation holds for the orbital entanglement data within the $\pi^\ast$-manifold, namely that they are the result of the departure from zero occupancy of the $\pi^\ast$-MOs, predominantly of the LUMO and LUMO+1.

Naively, one would expect that the single orbital entanglement of the PM-localized and atomic-like orbitals  should qualitatively show the same kind of deviation from that of the canonical MOs, with the latter being stronger for the atomic-like orbitals. However this is clearly not the case here. The single orbital entanglement of the PM-localized MOs is much more uniformly distributed than that of the canonical ones, and overall visibly lower. On the one hand, the uniformity of the entanglement originates from the near translation invariance of the PM-localized MOs. Moreover, the entanglement differs only slightly from uniformity when the MO is located at the edge of the carbon-carbon chain, where a boundary effect comes into play. On the other hand, the lower value of entanglement of the PM-localized MOs is rooted in their geometrical shape. Each MO is centered around two carbon atoms involved in a chemical $\pi$-bond, effectively masking the entanglement of the bond within each MO itself. Interestingly, from the orbital plots shown in the upper panels of Figures \ref{fig:C10H12_s1}\ and \ref{fig:C20H22_s1}, we conclude that the PM-localized MOs can still be identified as either bonding $\pi$- or anti-bonding $\pi^\ast$-MOs. Notably, the N-SSR entanglement of the PM-localized orbitals is almost zero compared to that in the canonical case, which is a clear indication that the PM-localized orbitals have a higher degree of zero seniority (close to either zero or double occupancy) than the canonical ones.

By contrast, the single orbital entanglement of the atomic-like orbitals is almost equally distributed. This finding is perhaps not surprising as the atomic-like orbitals  are almost identical up to translation and a possible phase change. Furthermore, the high amount of both P-SSR and N-SSR entanglement is a strong indicator of the prominent superposition in both the even and odd parity sectors. In stark contrast to the PM-localized orbitals, the degree of entanglement becomes substantially higher than that of the canonical MOs, reaching $96\%$ of the theoretical maximum of $2\ln 2$. To explain this finding, we recall that the atomic-like orbitals  are, by construction, a superposition of bonding and anti-bonding PM-localized orbitals. Such a rotation between matching $\pi$-$\pi^\ast$-MOs entails a release of the entanglement tucked away within the PM-localized orbitals, and becomes manifest in an entanglement between the atomic-like orbitals . As we shall see in Section \ref{subsubsec:orb-orb}, each atomic-like orbitals centered on one carbon atom has a pairwise entanglement with exactly only one other atomic-like orbital, localized around the second carbon center that is contributing to the same chemical bond.

\begin{figure*}[htb]
    \centering
    \includegraphics[scale=0.495]{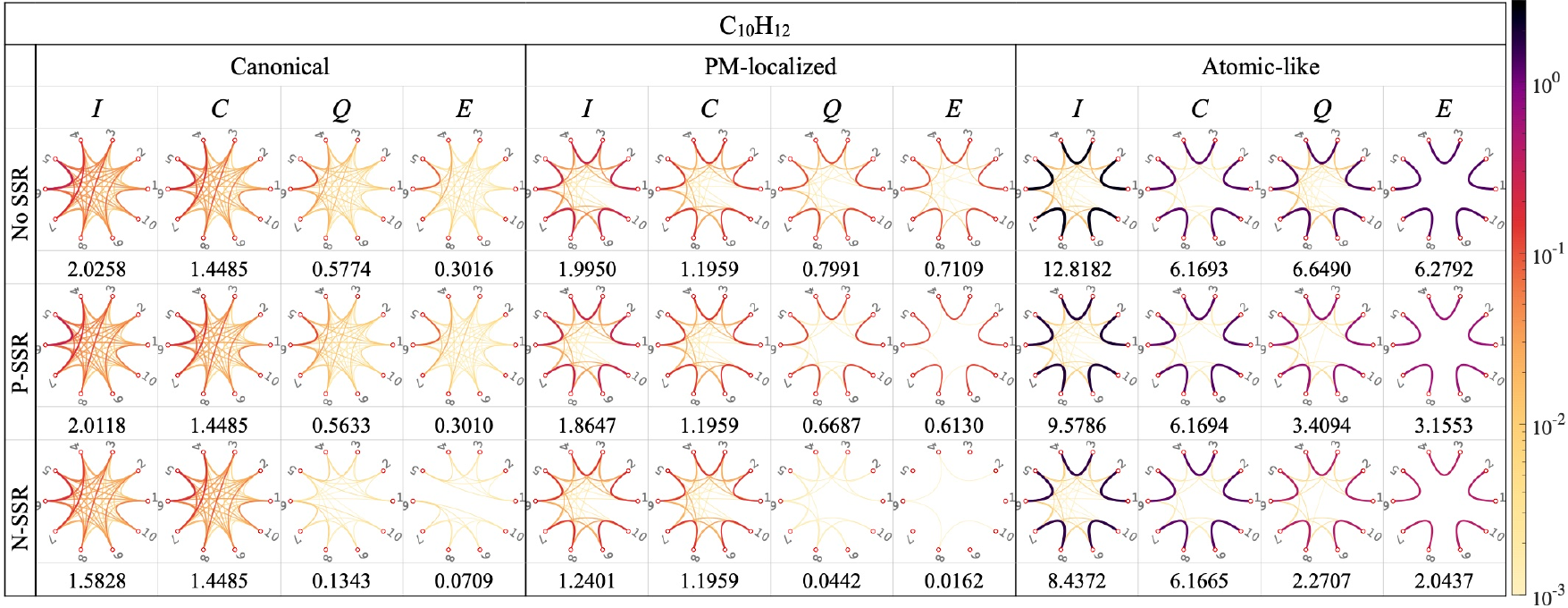}
    \caption{Pairwise orbital total correlation $I$, classical correlation $C$, quantum correlation $Q$, and entanglement $E$ in the CAS(10,10)-optimized electronic ground state of $\mathrm{C_{10}H_{12}}$\ in case of no SSR, P-SSR, and N-SSR. The orbital numbering follows the one given for the canonical, PM-localized, and atomic-like orbitals in the upper panel of Figure \ref{fig:C10H12_s1}. The corresponding pairwise correlation sum (see Eq.~\eqref{eqn:corrsum}) is given below each plot.}
    \label{fig:C10H12_s2}
\end{figure*}

\begin{figure*}[htb]
    \centering
    \includegraphics[scale=0.495]{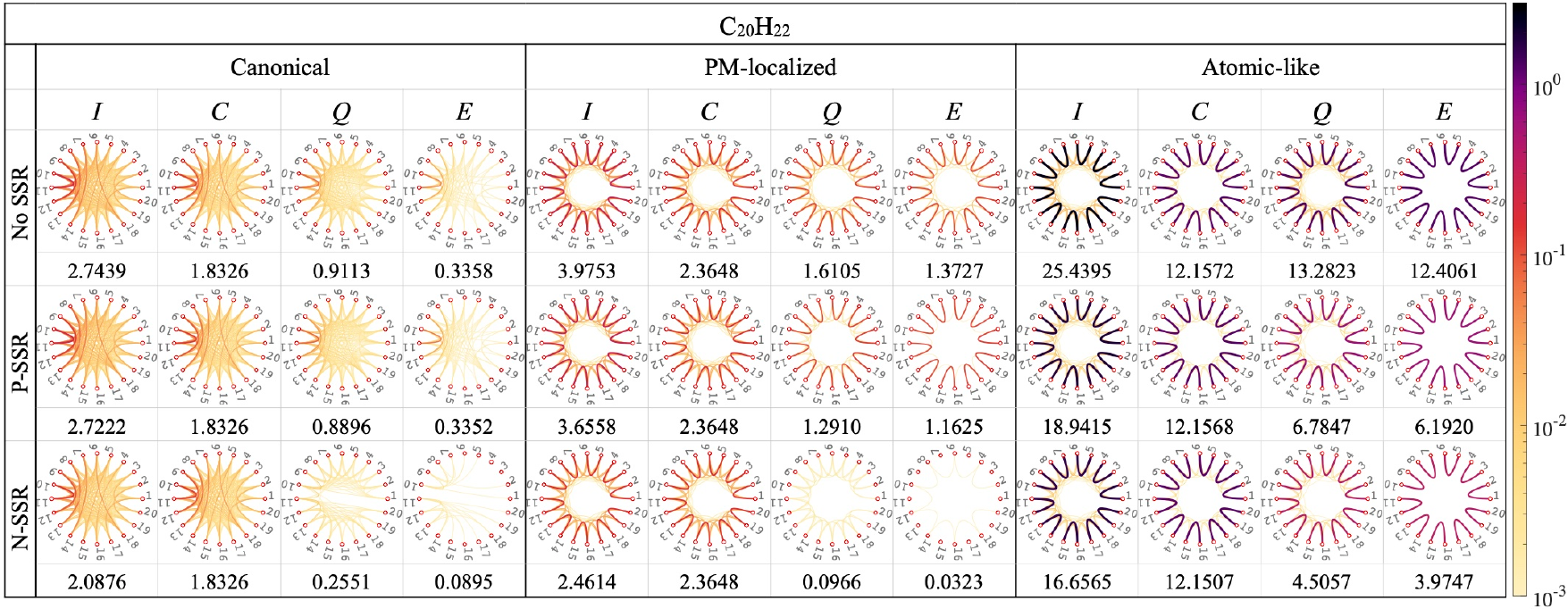}
    \caption{Pairwise orbital total correlation $I$, classical correlation $C$, quantum correlation $Q$, and entanglement $E$ in the CAS(20,20)-optimized electronic ground state of $\mathrm{C_{20}H_{22}}$\ in case of no SSR, P-SSR, and N-SSR. The orbital numbering follows the one given for the canonical, PM-localized, and atomic-like orbitals in the upper panel of Figure \ref{fig:C20H22_s1}. The corresponding pairwise correlation sum (see Eq.~\eqref{eqn:corrsum}) is given below each plot.}
    \label{fig:C20H22_s2}
\end{figure*}

\subsubsection{Orbital-Orbital Correlations} \label{subsubsec:orb-orb}

In this section, we analyze the orbital-orbital correlations (classical, quantum, entanglement) in the ground states of $\mathrm{C_{10}H_{12}}$ and $\mathrm{C_{20}H_{22}}$, as a continuation of the previous section which focused on the single orbital entanglement.
We consider the two-orbital reduced density states as the ``overall'' state, which are typically mixed, and serve as our point of departure to study the correlation between any two orbitals (within our correlation model CAS space).
In this scenario, the total correlation $I$ is no longer linearly related to the entanglement $E$, and the latter is therefore always smaller or equal to the quantum correlation $Q$.

In order to enable an unambiguous comparison of correlation strengths throughout our various choices of MO bases, we define the following quantities as the pairwise total correlation sum $I_\text{sum}$, pairwise classical correlation sum $C_\text{sum}$, pairwise quantum correlation sum $Q_\text{sum}$, and pairwise entanglement sum $E_\text{sum}$,
\begin{equation} \label{eqn:corrsum}
\begin{split}
    X_\text{sum}^{(\text{P, N})}(\{\phi_l\},|\Psi\rangle) &= \sum_{i<j} X(\rho_{ij}^{(\text{P, N})})
    \\
    X &= I, C, Q, E
\end{split}
\end{equation}
where $\rho_{ij}^{(\text{P, N})}$ is the (P, N-SSR compatible) reduced state of $|\Psi\rangle$ on the orbital $\phi_i$ and $\phi_j$ of the specified basis set $\{\phi_l\}$.

In Figure \ref{fig:C10H12_s2} and \ref{fig:C20H22_s2} we highlight the orbital-orbital total correlation $I$, classical correlation $C$, quantum correlation $Q$, and entanglement $E$ between the canonical, PM-localized, and atomic-like orbitals  in the ground states of $\mathrm{C_{10}H_{12}}$ and $\mathrm{C_{20}H_{22}}$, respectively. Moreover, the corresponding pairwise correlation sum is shown below each plot. We discuss in the following three major conclusions that can be drawn from the orbital-orbital correlation data.

We \textit{first} observe a primarily low total correlation $I$\ between either canonical or PM-localized MOs, whereas the degree of correlation between the atomic-like orbitals is strikingly higher,  exhibiting a six-fold increase in going from the PM-localized to the atomic-like orbitals basis. Simultaneously, the pairwise entanglement $E$\ reaches up to $91\%$ of $2\ln 2$ for both $\mathrm{C_{10}H_{12}}$ and $\mathrm{C_{20}H_{22}}$. Moreover, in the atomic-like orbitals basis, we find that the pairwise quantum correlation $Q$\ can be as large as $94\%$ of $2\ln 2$. We already saw in the case of $\mathrm{C_2H_4}$ that $100\%$ of maximal entanglement can never occur in an interacting molecule. Here however, the entanglement is further lowered due to the presence of orbital coupling. To see this we first notice that the two-orbital reduced states are now mixed, as a result of interaction between the two orbitals and the rest of the system. This degree of mixedness indicates that the two orbital system is also entangled with other orbitals, and hence naturally reduces the maximally achievable entanglement between them.

In addition to a comparison of absolute correlation data, it is instructive to consider the relative contributions of quantum and classical correlation to the total correlation, focusing first on the case without SSRs.
In the canonical MO basis, a larger portion of the total correlation is classical rather than quantum in nature. For example, the pairwise quantum correlation sum $Q_{\rm sum}$ in the ground state of $\mathrm{C_{10}H_{12}}$ is only $29\%$ of the pairwise total correlation sum $I_{\rm sum}$, and, similarly, $33\%$ for $\mathrm{C_{20}H_{22}}$. As we move to the PM-localized MO basis, though the overall total correlation does not increase, the relative contribution of quantum correlation $Q$ rises to $40\%$ and $41\%$ for $\mathrm{C_{10}H_{12}}$ and $\mathrm{C_{20}H_{22}}$ respectively. This effect becomes even more apparent in the atomic-like orbitals basis, where the fraction of quantum correlation ($> 52\%$) surpasses that of the classical correlation for both molecules. Furthermore, in passing from a canonical as well as PM-localized to an atomic-like orbitals basis, we not only observe an increase of the percentage of quantum correlation comprised in the total correlation, but also encounter a significant increase of the share of entanglement $E$\ in the quantum correlation. The latter increases from $37\%$ and $85\%$ to up to $93\%$ for $\mathrm{C_{20}H_{22}}$, as the orbitals are becoming more and more localized. \textit{Secondly}, besides the effects on the importance of quantum correlation and entanglement, (almost fully) localizing the MOs introduces a distinct pairing structure. Among the canonical MOs, we do not find any obvious pairing structure except for those MOs located around the Fermi level (HOMO--LUMO, HOMO-1--LUMO-1) which also exhibit the largest pairwise correlations. Moving to the PM-localized MOs, a clear pairing structure emerges with pairs of MOs grouped together by relatively strong (total) pair-wise correlation and entanglement. Nonetheless, such a pair-wise correlation is still too weak to fully describe a chemical bond, compared to the maximal entanglement we found in Subsection \ref{subsec:Hbond}. The pattern observed in this case stems from the fact that the two pairing MOs are the respective bonding $\pi$- and anti-bonding $\pi^\ast$-orbitals located primarily on the same two carbon atoms, an effect similar to the correlation between the canonical and PM-localized MOs in $\mathrm{C_2H_4}$ in Section \ref{subsec:ethylene}. In a mean-field picture, all the local bonding orbitals are doubly occupied while the anti-bonding ones remain empty. By contrast, in a correlated picture electron-electron interaction introduces a finite occupation in the latter while simultaneously reducing the double-occupancy occupation of the former, thus creating a \textit{weak} pair-wise correlation between the local bonding and anti-bonding orbitals. Considering next the atomic-like orbitals basis, the overall picture changes strikingly. The pronounced pairing structure of the correlation and entanglement data between MOs located on neighboring carbon atoms is an order of magnitude stronger than that of the PM-localized MOs. The pairwise entanglement $E$ becomes even so strong that a so-called monogamy effect results, namely each atomic-like orbitals is \textit{only} entangled to one other MO with which a bond is formed, in sharp contrast to the weak correlation background (yellow connecting lines) in the quantum correlation plots where MOs from different pairs are still correlated. This clear distinction indicates that (pair-wise) entanglement may be a more appropriate quantity to describe chemical bonds than quantum correlation would be.

\textit{Finally}, we would like to highlight the effect of SSRs and their implication on the resulting complexity of the ground state wave function. For the canonical and PM-localized MO basis, P-SSR hardly changes the pairwise entanglement, whereas N-SSR suppresses almost all of it. This suggests that the dominating configurations in the total wave function are those with double or zero occupancy on the respective orbitals. In particular, the leading configurations are those where the bonding $\pi$-orbitals are doubly occupied rather than the anti-bonding $\pi^\ast$\ ones, giving rise to a weak pairing structure that could not survive in the presence of N-SSR. In other words, within the two-electron Hilbert space of two pairing orbitals (one bonding $\pi_i$ and anti-bonding $\pi_j^\ast$), the leading configuration is simply $|\!\!\uparrow\!\downarrow\rangle_{\pi_i}\otimes|0\rangle_{\pi_j^\ast}$. Turning instead to atomic-like orbitals $\phi_i$ and $\phi_j$ that form a chemical bond, a far more complicated structure of the ground state emerges in this basis. For a given pair $i,j$\ of strongly entangled orbitals, none of the four configurations with fixed local occupation numbers
\begin{equation}
\begin{split}
    &|\!\uparrow\!\downarrow\rangle_{\phi_i} \otimes |0\rangle_{\phi_j}, \quad |0\rangle_{\phi_i} \otimes |\!\uparrow\!\downarrow\rangle_{\phi_j},
    \\
    &|\!\uparrow\rangle_{\phi_i} \otimes |\!\downarrow\rangle_{\phi_j}, \quad \: |\!\downarrow\rangle_{\phi_i}\otimes|\!\uparrow\rangle_{\phi_j},
\end{split}
\end{equation}
in the two-electron Hilbert space is particularly dominating, as the entanglement is nearly at its maximum (if we only consider the two-electron subspace). This is also the reason why we encounter a strong P-SSR and N-SSR entanglement between the atomic-like orbitals . The drastically increased number of configurations with appreciable non-zero weight in the total wave function leads to a strongly, and \emph{statically} correlated ground state description. Hence, even though an atomic-like orbitals basis is most suitable for a genuine description of a chemical bonding, in particular in terms of the pair-wise entanglement $E$, this basis, compared to the canonical and PM-localized counterparts, introduces a strong correlation structure in the electronic molecular ground state wave function. More importantly, though, the high degree of entanglement and quantum correlation even under P-SSR and N-SSR makes it a prime candidate in quantum computing applications.

To briefly summarize, we have successfully applied our orbital localization scheme first time to systems of realistic sizes, harnessing inspirations from both analytic example in Subsection \ref{subsec:Hbond} and smaller concrete system, namely the ethylene molecule in Subsection \ref{subsec:ethylene}. Astonishingly, we were able to both locate and measure the intensity of the $\pi$-bonds using the monogamous entanglement between the fully-localized atomic-like orbitals, essentially breaking down the chain systems into units of ethylenes. The near maximal entanglement level we found in all the carbon chain systems established again that a single bond correspond to the entanglement value of $E=2\ln 2$ between the two contributing atomic-like orbitals, though marginally lessened by electronic interaction. On one hand, our comprehensive QI framework offer surgical analysis of correlation and entanglement; on the other hand, the fully-localized atomic-like orbitals are the chemically meaningful subject of study. Together, they reveal a deep connection between bond order and the entanglement between the involved nuclear centers represented by the respective atomic-like orbitals, and offer a quantitative perspective to the valence bonding theory.

\subsection{Benzene} \label{subsec:benzene}

\begin{figure*}[htb]
    \centering
    \includegraphics[width=0.6\textwidth]{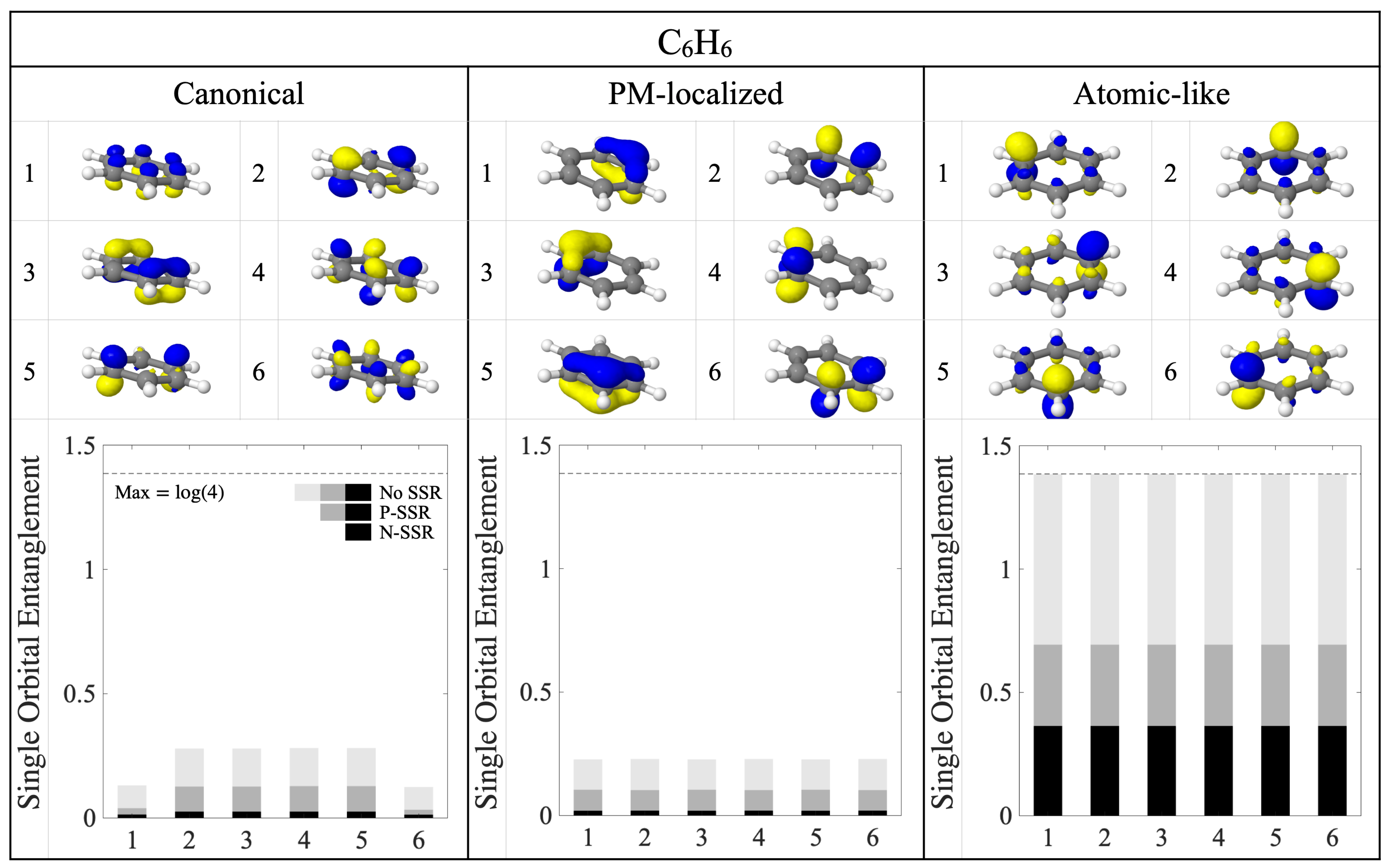}
    \caption{Single orbital entanglement in the CAS(6,6)-optimized electronic ground state of benzene. The orbital numbering on the $x$-axis in the lower panel follows the one given for the canonical, PM-localized, and atomic-like orbitals in the upper panel (plotted with an isosurface value of 0.05 for the canonical and PM-localized MOs, and 0.1 for the atomic-like orbitals ). The color code for the single orbital entanglement data is as follows: no SSR (all color), P-SSR (black and dark grey), and N-SSR (black).}
    \label{fig:C6H6_s1}
\end{figure*}

\begin{figure*}[htb]
    \centering
    \includegraphics[scale=0.495]{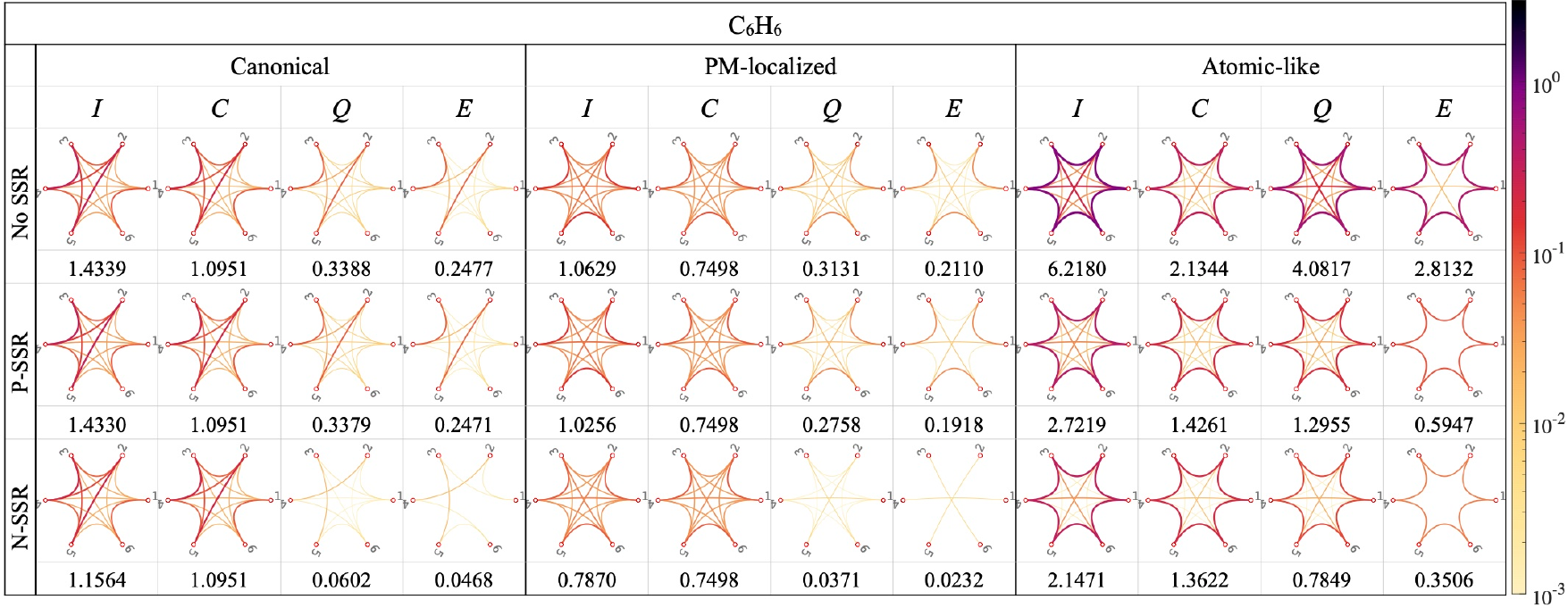}
    \caption{Pairwise orbital total correlation $I$, classical correlation $C$, quantum correlation $Q$, and entanglement $E$ in the CAS(6,6)-optimized electronic ground state of benzene in case of no SSR, P-SSR, and N-SSR. The orbital numbering follows the one given for the canonical, PM-localized, and atomic-like orbitals in the upper panel of Figure \ref{fig:C6H6_s1}. The corresponding pairwise correlation sum (see Eq.~\eqref{eqn:corrsum}) is given below each plot.}
    \label{fig:C6H6_s2}
\end{figure*}

Having discussed the linear, $\pi$-conjugated systems, we will in this section focus on the correlation pattern of the ground state of \emph{the} prototypical, cyclic, $\pi$-conjugated aromatic molecule, namely benzene ($\mathrm{C_6H_6}$). 

To ease comparison, we perform the same correlation analysis as in Section \ref{subsec:polyene}, by making use of the same three distinct sets of MO  bases: canonical, PM-localized, and atomic-like orbitals , whose isosurfaces are shown in Figure \ref{fig:C6H6_s1}. Before embarking on the correlation analysis, a comment is in order on how the atomic-like orbitals basis can be obtained for benzene.
Compared to the other $\pi$-conjugated molecular systems, an atomic-like orbitals basis for benzene requires a generalized localization scheme. To explain this fact, we first point out that, for the polyenes, the PM-localized orbitals already indicated how atomic-like orbitals  can be obtained. As we can see in Figure \ref{fig:C10H12_s1} and \ref{fig:C20H22_s1}, the PM-localized orbitals of $\mathrm{C_{10}H_{12}}$ and $\mathrm{C_{20}H_{22}}$ are of bonding and anti-bonding $\pi$-type, respectively, stretching primarily across two carbon centers. Making then use of the insight from our analytic example in Section \ref{subsec:Hbond}, it is straightforward to see that the respective atomic-like orbitals\ result from a unitary rotation with angle $\theta$\ of the bonding and anti-bonding orbital pairs located on the same carbon centers by $\theta=\frac{\pi}{4}$. By contrast, such a pairing structure no longer emerges for the PM-localized basis in case of benzene. First, the PM-localized orbitals span over more than two carbon centers, due to the absence of a fixed local bonding region. Second, there is no obvious way of rotating any two orbitals which could give rise to atomic-like ones, solely based on geometrical considerations. Hence, we propose in the following a systematic way that leads to an atomic-like orbitals basis starting from the canonical rather than the PM-localized MO basis.

The canonical $\pi$-MOs of benzene (denoted as $\phi_i$'s) are linear combinations of atomic orbitals (LCAO)
\begin{equation}
    \tilde{\phi}_i = \sum_j \mathbf{U}_{ij} \alpha_j. \label{eqn:atomic_to_HF}
\end{equation}
where $\alpha_i$'s are the atomic orbitals and $\mathbf{U}$ a unitary matrix. Because the atomic orbitals have finite overlap with each other, the LCAO's need to be further normalized
\begin{eqnarray}
\phi_i = \mathbf{\Lambda}_{ij} \tilde{\phi}_j = (\mathbf{\Lambda}\mathbf{U})_{ij} \alpha_j = \frac{\tilde{\phi}_i}{\|\tilde{\phi_i}\|}.
\end{eqnarray}
This last normalization transformation $\mathbf{\Lambda}$ makes the mapping $\mathbf{\Lambda} \mathbf{U}$ from the atomic orbitals to the canonical orbitals non-unitary. Hence, we may ask ourselves what would happen if we transform the canonical orbitals by $\mathbf{U}^{-1}$? The resulting orbitals are still orthogonal of course, since $\mathbf{U}^{-1}$ is unitary. To assess the locality of these orbitals, we first consider an extreme example. Suppose that we have a molecule with internuclear distances such that all of its atomic orbitals (at different atomic centers) have vanishingly small overlap with each other. This entails that the LCAO's in Eq.~\eqref{eqn:atomic_to_HF} are already normalized ($\mathbf{\Lambda}=\openone$). Then the inverse transform
\begin{equation}
    \tilde{\alpha}_i = \sum_j (\mathbf{U}^{-1})_{ij} \phi_j = \sum_j (\mathbf{U}^{-1}\mathbf{\Lambda}\mathbf{U})_{ij} \alpha_j
\end{equation}
would only give back the original atomic orbitals $\tilde{\alpha_i} = \sum_j(\mathbf{U}^{-1}\mathbf{U})_{ij} \alpha_j = \alpha_i$. When the atomic overlap is finite, $\mathbf{\Lambda}$ then deviates from the identity map, and so does $\mathbf{U}^{-1}\mathbf{\Lambda}\mathbf{U}$. However, as long as the atomic overlap is not exceedingly large\footnote{For example, straightforward calculation would show for the case of two identical atomic orbitals with overlap of 0.75, each resulting atomic-like orbital is still dominated by one atomic orbital with 69\% relative weight.}, $\mathbf{\Lambda}$\ is close to an identity map, and each orbital $\overline{\alpha}_i$ will have a dominating contribution from the atomic orbital $\alpha_i$ and, simultaneously only small weights from the remaining ones, thus making the new \textit{molecular} orbitals $\tilde{\alpha}_i$ atomic-like.
Remarkably, our simple scheme does not refer to any cost function or require numerical optimization on a particular software platform. Furthermore, the atomic-like orbitals  play the same role as the atomic orbitals in molecular orbital theory, in the sense that the same linear combinations of the atomic-like orbitals  as those of the atomic ones reproduce the canonical MOs, up to a normalization factor.

Consequently, we applied for benzene the above localization scheme starting out from the canonical MO basis. As can be seen from Figure \ref{fig:C6H6_s1}, each atomic-like orbitals has a large contribution from a $p_z$-shaped \textit{atomic} orbital on one carbon center, and only small weights from the remaining $p_z$-shaped orbitals on the other carbon centers. Moreover, all  atomic-like orbitals  are identical up to translations by an integer multiple of the lattice constant along the benzene ring. The localization scheme therefore successfully preserves the original shape of the atomic orbitals, while maintaining the orthogonality of the canonical ones.

In Figures \ref{fig:C6H6_s1} and \ref{fig:C6H6_s2}, respectively, we present the single-orbital and orbital-orbital correlation results for all three sets of MO bases. The correlation patterns within the canonical and PM-localized MO basis mainly resemble those of the linear polyenes discussed in the previous Subsection. In more detail, we first observe an overall low single orbital and orbital-orbital entanglement. Secondly, in the case without SSR most of the orbital-orbital correlation is again classical in nature with a share of $76\%$ and $71\%$, respectively, of the orbital-orbital total correlation for the two MO bases. Thirdly, the distribution pattern of the superselected single orbital entanglement is also in alignment with previous findings. The PM-localized orbitals display the lowest level of N-SSR entanglement due to their higher degree of zero seniority, whereas the atomic-like orbitals exhibit high P-SSR and N-SSR entanglement, indicating strong superposition in both even and odd parity sectors.

Similar to the situation encountered for polyene systems, the correlation structure in the atomic-like orbitals basis becomes much richer. Considering first the single-orbital entanglement without SSR, we find that every atomic-like orbitals is maximally entangled (with value $2\ln 2$) with the rest of the system, as can be seen from Figure \ref{fig:C6H6_s1}. This strong entanglement is missing at the orbital-orbital level, though. A closer inspection of Figure \ref{fig:C6H6_s2} reveals that the maximal pairwise entanglement $E$\ is considerably weaker than the one between the atomic-like orbitals  in the polyenes. Compared to $92\%$ of $2\ln 2$ for the latter, the maximum value of $E$\ in benzene amounts to merely  $33\%$ of $2\ln 2$. The origin of this discrepancy can be explained as follows. In sharp contrast to the pairing structure of the entanglement between the atomic-like orbitals  in the polyenes, each atomic-like orbital on the benzene ring is equally entangled to \textit{both} of its neighbors. This ``left-right pairing" is rooted in the underlying symmetry of the molecular Hamiltonian and closely resembles the simplified ``left-right overlap"\ model of the H\"uckel Hamiltonian for benzene\cite{huck31a}. Hence, the unique electronic ground state within our minimal $\pi$-$\pi^\ast$\ CAS(6,6) active orbital space enjoys the same symmetry as the molecule itself, and is invariant under six-fold rotation and reflection about the mirror planes. A simplified, polyene-like pairing structure which would gives rise to a $\pi$-type bond involving only two neighboring carbon centers is therefore suppressed by the molecular symmetry. Although, even if we are to sum up the orbital-orbital entanglement between one orbital and both of its neighbors, $33\%$ of the single orbital entanglement would not be accounted for from this partially summed \textit{bipartite} orbital-orbital entanglement sum. As a result, we ascribe the missing part of the entanglement in this $\pi$-conjugated, aromatic molecule to genuine \textit{multipartite} entanglement. Interestingly, for the cyclopropyl cation, the smallest 2-$\pi$-electron cyclic aromatic molecule (results not shown here), we do \textit{not}\ encounter signs of entanglement beyond the orbital-orbital picture between the three carbon centers in the odd-numbered ring skeleton. Thus, whether multipartite entanglement is a distinct feature of $\pi$-conjugated, aromatic molecular systems with an \textit{even} number of atomic centers contributing to the $\pi$-conjugation clearly deserves further investigation. While it goes beyond the scope of our current work, it will be subject of a future study.

\section{Summary and Conclusion}\label{sec:concl}

In this work we established a comprehensive quantum information framework for electronic structure analysis with the aim to foster synergies of the quantum information (QI) and quantum chemistry (QC) communities within the
second quantum revolution. This framework enabled us to dissect and compare quantum and classical effects in molecular systems, which closes an apparent gap in the current literature. This was made possible thanks to our unifying geometric picture of quantum states, which facilitated a unique definition of these correlation quantities on the same footing.

Under this framework, opened then up the possibility to quantify quantum correlation within molecules as a resource for information processing tasks. Still, since the amount of quantum correlation depends on the chosen orbital basis, we strove to maximize the resourcefulness of molecular orbitals through an orbital optimization scheme. The key idea of our scheme was to localize each molecular orbital to one atomic center only. This constituted a significant step beyond ``traditional" localization schemes (e.g., Pipek-Mezey (PM)) which typically yield final orbitals spanning over more than one nucleus. Combining both new essential ingredients --- the comprehensive QI framework and our fully-localized orbitals --- we arrived at two key results: (i) maximal resourcefulness of molecular systems is realized by fully-localized orbitals. (ii) A single covalent bond is quantitatively best rationalized by the maximal orbital-orbital entanglement, that is $2\ln2$.

To showcase these two crucial insights, we systematically compared various correlation quantities with respect to three important bases (canonical, PM-localized and fully-localized), in the ground states of $\pi$-conjugated chain systems ($\mathrm{C_2H_4}$, $\mathrm{C_{10}H_{12}}$ and $\mathrm{C_{20}H_{22}}$). We observed almost maximal entanglement between the fully-localized orbitals, which is 9 and 37 times of the entanglement found in the PM-localized and canonical basis, respectively (comparing $E_\text{sum}$ in $\mathrm{C_{20}H_{22}}$). This drastic difference unequivocally shows that a  fully-localized orbital basis is not only by far a superior reference but also reveals the abundant quantum resource inside these systems. Concomitantly, our analysis enabled us to associate \textit{quantitatively} the existence of a chemical bond with the entanglement between fully-localized orbitals: each fully-localized orbital is only, \textit{and} maximally, entangled with one other, together with which a bond is formed. Our entanglement analysis thus constitutes an important milestone in turning valence bond theory into a fully quantitative machinery.

To thoroughly further quantify the relation between chemical bonding and orbital entanglement, we also studied the prototypical aromatic molecule benzene. Unlike the polyene systems, the benzene ground state does not exhibit a conjugated $\pi$-bond structure. This is exactly reflected in our entanglement analysis: the entanglement in benzene is uniformly distributed, without an alternating pattern of strong and weak orbital-orbital entanglement. Yet, each fully-localized orbital was found to be maximally entangled with the complementary subsystem. The fact that this entanglement is not fully reflected in the orbital-orbital entanglement, confirms the commonly accepted picture that the $\pi$-bonds in benzene are shared within the entire ring rather than locally.

Finally, we would like to envisage a few promising directions sparked by the new framework we put forward in this work. Firstly, our quantitative rationalization of chemical bonding through entanglement analysis can be extended beyond $\pi$-bonds, such as to also encompass $\sigma$-type hybridized orbitals. The latter will therefore allow us to consider the vast majority of local bonds that dominate the chemical bonding of main group compounds. Secondly, we would like to strengthen the quantitative connection between orbital entanglement and highly delocalized, multicenter bonds, by exploring multipartite entanglement as a type of bonding descriptor, that is, for example, as an indicator for aromaticity. In conclusion, after showcasing versatile pathways to genuinely identify a potentially ample supply of quantum resource in chemical bonds, we look forward to spark an intense collaboration between the QI and QC communities in order to both harvest and further exploit these resources. This new prospect has the potential to bring about a paradigm-shifting change to modern chemistry: molecules are not only mere reactants in chemical processes but also potent sources of entanglement and quantum correlations, the required resources for information processing tasks in the age of quantum technologies.

\begin{acknowledgements}
We thank S.\hspace{0.5mm}Mardazad for helpful discussions.
We gratefully acknowledge K.\hspace{0.5mm}S.\hspace{0.5mm}Min for helpful comments on the manuscript.
We acknowledge financial support from the Deutsche Forschungsgemeinschaft (DFG, German Research Foundation), Grant SCHI 1476/1-1 (L.D., C.S.), the Munich Center for Quantum Science and Technology (L.D., C.S.) and the NKFIH through the Quantum Information National Laboratory of Hungary program, and Grants No. K124152, FK135220, K124351 (Z.Z.). The project/research is also part of the Munich Quantum Valley, which is supported by the Bavarian state government with funds from the Hightech Agenda Bayern Plus.
\end{acknowledgements}

\bigskip

\appendix
\setcounter{table}{0}
\renewcommand{\thetable}{\Alph{section}.\arabic{table}}

\section{Correlation Sum Rule} \label{app:corr_sum}

In this section we will show that the quantum and classical correlation of $\rho$\ sum up to the total correlation of $\rho$ if its closest product state $\pi_\rho$ and its closest classical state $\chi_\rho$ are simultaneously diagonalized.

We first observe that
\begin{equation}\label{eqn:diff}
\begin{split}
    I(\rho)-&Q(\rho)-C(\rho) \qquad \qquad
        \\
        &= S(\rho_A) + S(\rho_B) - S(\chi_A) - S(\chi_B),
\end{split}
\end{equation}
where $\chi_{A,B}$ are the reduced states of $\chi_\rho$, respectively. We can then show that the spectrum of $\rho_A$ and that of $\chi_A$ coincide. Let us denote the eigenstate of $\chi_\rho$ as $\{|i\rangle\!\otimes\!|j\rangle\}$. From this it follows that the eigenvalues of $\rho_A$ are precisely its diagonal entries in this basis
\begin{equation}
    \lambda_i = (\rho_A)_{ii} = \langle i| \rho_A |i\rangle = \sum_j \langle i| \!\otimes\! \langle j | \rho |i \rangle \!\otimes\! |j\rangle.
\end{equation}
Moreover, the eigenvalues of $\chi_A$ are given by
\begin{equation}
    \begin{split}
        \mu_i &= \langle i | \Tr_B[\chi_\rho] |i\rangle
        \\
        &=\langle i | \Tr_B\left[\sum_{kj}|k\rangle\langle k|\!\otimes\! |j\rangle\langle j| \rho |k\rangle\langle k|\!\otimes\! |j\rangle\langle j|\right] |i\rangle
        \\
        &= \sum_j \langle i| \!\otimes\! \langle j | \rho |i \rangle \!\otimes\! |j\rangle = \lambda_i.
    \end{split}
\end{equation}
Using similar considerations, the spectrum of $\rho_B$ and $\chi_B$ also coincide. As a result, the right-hand side of Eq.\eqref{eqn:diff} must vanish, leading to the sum rule
\begin{equation}
    I(\rho) = Q(\rho) + C(\rho).
\end{equation}
This allows us to dissect the total correlation exactly into quantum and classical correlation contributions. When $\rho_{A,B}$ is not simultaneously diagonalized as $\chi_{A,B}$, its diagonal entries still coincide with those of $\chi_{A,B}$, namely the elements of $\Vec{\mu}$. From this, it follows that the spectrum of $\rho_{A,B}$\ given by $\Vec{\lambda}$, majorizes $\Vec{\mu}$, leading to the inequality $S(\rho_{A,B}) > S(\chi_{A,B})$. Hence, a general relation between the total, quantum, and classical correlation reads as
\begin{eqnarray}
I(\rho) \geq Q(\rho) + C(\rho).
\end{eqnarray}

\section{Reference structures}

\begin{table}[H]
    \caption{XYZ coordinates (in \AA)\ of the benzene structure as taken from Ref.~[\onlinecite{erik20a}].}
    \label{tab:benzene:coord}
    \centering
     \begin{tabular}{lrrr}\\
C  & 0.0000000000 &      1.3967920000 &      0.0000000000 \\
C  & 0.0000000000 &     -1.3967920000 &      0.0000000000 \\
C  & 1.2096570000 &      0.6983960000 &      0.0000000000 \\
C  &-1.2096570000 &     -0.6983960000 &      0.0000000000 \\
C  &-1.2096570000 &      0.6983960000 &      0.0000000000 \\
C  & 1.2096570000 &     -0.6983960000 &      0.0000000000 \\
H  & 0.0000000000 &      2.4842120000 &      0.0000000000 \\
H  & 2.1513900000 &      1.2421060000 &      0.0000000000 \\
H  &-2.1513900000 &     -1.2421060000 &      0.0000000000 \\
H  &-2.1513900000 &      1.2421060000 &      0.0000000000 \\
H  & 2.1513900000 &     -1.2421060000 &      0.0000000000 \\
H  & 0.0000000000 &     -2.4842120000 &      0.0000000000
    \end{tabular}
\end{table}

\begin{table}[H]
    \caption{XYZ coordinates (in \AA)\ of the ethylene structure as taken from Ref.~[\onlinecite{herz66a}].}
    \label{tab:ethylene:coord}
    \centering
     \begin{tabular}{lrrr}\\
C &       0.6695000000 &      0.0000000000  &     0.0000000000\\
C &      -0.6695000000 &      0.0000000000  &     0.0000000000\\
H &       1.2321000000 &      0.9289000000  &     0.0000000000\\
H &       1.2321000000 &     -0.9289000000  &     0.0000000000\\
H &      -1.2321000000 &      0.9289000000  &     0.0000000000\\
H &      -1.2321000000 &     -0.9289000000  &     0.0000000000
    \end{tabular}
\end{table}

\begin{table}[H]
    \caption{XYZ coordinates (in \AA)\ of the decapentaene  structure as taken from Ref.~[\onlinecite{huwe15a}].}
    \label{tab:decapentaene:coord}
    \centering
     \begin{tabular}{lrrr}\\
C &      -5.5708190000   &   -0.2177430000  &     0.0000000000\\
H &      -5.6461840000   &   -1.2984100000  &     0.0000000000\\
H &      -6.4957320000   &    0.3422470000  &     0.0000000000\\
C &      -4.3734900000   &    0.4059890000  &     0.0000000000\\
H &      -4.3479600000   &    1.4905820000  &     0.0000000000\\
C &      -3.0858560000   &   -0.2773930000  &     0.0000000000\\
H &      -3.1060170000   &   -1.3624430000  &     0.0000000000\\
C &      -1.8851500000   &    0.3575050000  &     0.0000000000\\
H &      -1.8687940000   &    1.4426720000  &     0.0000000000\\
C &      -0.6004210000   &   -0.3187700000  &     0.0000000000\\
H &      -0.6156070000   &   -1.4038340000  &     0.0000000000\\
C &       0.6004230000   &    0.3187720000  &     0.0000000000\\
H &       0.6156100000   &    1.4038370000  &     0.0000000000\\
C &       1.8851510000   &   -0.3575040000  &     0.0000000000\\
H &       1.8687910000   &   -1.4426710000  &     0.0000000000\\
C &       3.0858590000   &    0.2773880000  &     0.0000000000\\
H &       3.1060260000   &    1.3624380000  &     0.0000000000\\
C &       4.3734890000   &   -0.4060020000  &     0.0000000000\\
H &       4.3479520000   &   -1.4905940000  &     0.0000000000\\
C &       5.5708250000   &    0.2177170000  &     0.0000000000\\
H &       5.6462120000   &    1.2983820000  &     0.0000000000\\
H &       6.4957270000   &   -0.3422920000  &     0.0000000000
    \end{tabular}
\end{table}

\begin{table}[H]
    \caption{XYZ coordinates (in \AA)\ of the eicosadecaene  structure as taken from Ref.~[\onlinecite{huwe15a}].}
    \label{tab:eicosadecaene:coord}
    \centering
     \begin{tabular}{lrrr}\\
C  &    -11.7770020000   &   -0.2649560000  &     0.0000000000\\
H  &    -11.8388260000   &   -1.3466900000  &     0.0000000000\\
H  &    -12.7091350000   &    0.2833280000  &     0.0000000000\\
C  &    -10.5885700000   &    0.3732840000  &     0.0000000000\\
H  &    -10.5763580000   &    1.4583190000  &     0.0000000000\\
C  &     -9.2921000000   &   -0.2943950000  &     0.0000000000\\
H  &     -9.2992590000   &   -1.3798130000  &     0.0000000000\\
C  &     -8.1003310000   &    0.3546870000  &     0.0000000000\\
H  &     -8.0971520000   &    1.4401960000  &     0.0000000000\\
C  &     -6.8073860000   &   -0.3058800000  &     0.0000000000\\
H  &     -6.8092390000   &   -1.3912330000  &     0.0000000000\\
C  &     -5.6148640000   &    0.3461770000  &     0.0000000000\\
H  &     -5.6138660000   &    1.4316090000  &     0.0000000000\\
C  &     -4.3232520000   &   -0.3125970000  &     0.0000000000\\
H  &     -4.3240920000   &   -1.3979910000  &     0.0000000000\\
C  &     -3.1299320000   &    0.3397370000  &     0.0000000000\\
H  &     -3.1292390000   &    1.4251560000  &     0.0000000000\\
C  &     -1.8391520000   &   -0.3189270000  &     0.0000000000\\
H  &     -1.8400330000   &   -1.4043370000  &     0.0000000000\\
C  &     -0.6452140000   &    0.3330280000  &     0.0000000000\\
H  &     -0.6441700000   &    1.4184400000  &     0.0000000000\\
C  &      0.6450570000   &   -0.3261670000  &     0.0000000000\\
H  &      0.6436700000   &   -1.4115880000  &     0.0000000000\\
C  &      1.8393950000   &    0.3251020000  &     0.0000000000\\
H  &      1.8411290000   &    1.4105080000  &     0.0000000000\\
C  &      3.1294910000   &   -0.3349600000  &     0.0000000000\\
H  &      3.1273120000   &   -1.4203910000  &     0.0000000000\\
C  &      4.3239310000   &    0.3154210000  &     0.0000000000\\
H  &      4.3266930000   &    1.4008120000  &     0.0000000000\\
C  &      5.6142480000   &   -0.3459560000  &     0.0000000000\\
H  &      5.6108230000   &   -1.4313950000  &     0.0000000000\\
C  &      6.8084330000   &    0.3031450000  &     0.0000000000\\
H  &      6.8130470000   &    1.3884900000  &     0.0000000000\\
C  &      8.0995750000   &   -0.3609920000  &     0.0000000000\\
H  &      8.0932010000   &   -1.4464930000  &     0.0000000000\\
C  &      9.2933930000   &    0.2843740000  &     0.0000000000\\
H  &      9.3039450000   &    1.3697610000  &     0.0000000000\\
C  &     10.5875730000   &   -0.3877450000  &     0.0000000000\\
H  &     10.5714070000   &   -1.4727330000  &     0.0000000000\\
C  &     11.7783810000   &    0.2460120000  &     0.0000000000\\
H  &     11.8445280000   &    1.3274870000  &     0.0000000000\\
H  &     12.7083250000   &   -0.3059840000  &     0.0000000000
    \end{tabular}
\end{table}

\section{Atomic-like Orbitals in Benzene} \label{app:benzene}

The pre-normalized canonical Hartree-Fock orbitals $\tilde{\phi}_i$'s of the benzene ring are related to the $p_z$ orbitals $\alpha_i$'s by
\begin{equation}
    \tilde{\phi}_i = \mathbf{U}_{ij} \alpha_j,
\end{equation}
where
\begin{eqnarray}
\mathbf{U} = \begin{pmatrix}
\frac{1}{\sqrt{6}} & \frac{1}{\sqrt{6}} & \frac{1}{\sqrt{6}} & \frac{1}{\sqrt{6}} & \frac{1}{\sqrt{6}} & \frac{1}{\sqrt{6}}
\\
\rule{0pt}{2.6ex}\rule[-1.2ex]{0pt}{0pt}
-\frac{1}{\sqrt{12}} & - \frac{1}{\sqrt{3}} & -\frac{1}{\sqrt{12}} & \frac{1}{\sqrt{12}} & \frac{1}{\sqrt{3}} & \frac{1}{\sqrt{12}}
\\
\rule{0pt}{2.6ex}\rule[-1.2ex]{0pt}{0pt}
\frac{1}{2} & 0 & -\frac{1}{2} & -\frac{1}{2} & 0 & \frac{1}{2}
\\
\rule{0pt}{2.6ex}\rule[-1.2ex]{0pt}{0pt}
-\frac{1}{2} & 0 & \frac{1}{2} & - \frac{1}{2} & 0 & \frac{1}{2}
\\
\rule{0pt}{2.6ex}\rule[-1.2ex]{0pt}{0pt}
-\frac{1}{\sqrt{12}} &  \frac{1}{\sqrt{3}} & -\frac{1}{\sqrt{12}} & -\frac{1}{\sqrt{12}} & \frac{1}{\sqrt{3}} & -\frac{1}{\sqrt{12}}
\\
\rule{0pt}{2.6ex}\rule[-1.2ex]{0pt}{0pt}
\frac{1}{\sqrt{6}} & -\frac{1}{\sqrt{6}} & \frac{1}{\sqrt{6}} & -\frac{1}{\sqrt{6}} & \frac{1}{\sqrt{6}} & -\frac{1}{\sqrt{6}}
\end{pmatrix}.
\end{eqnarray}

The atomic-like orbitals $\tilde{\alpha}_i$ proposed by us is then obtained by applying the inverse $\mathbf{U}^{-1}$ to the \emph{normalized} canonical Hartree-Fock orbitals $\phi_i = \tilde{\phi}_i/\|\tilde{\phi}_i\|$
\begin{equation}
    \tilde{\alpha}_i = \mathbf{U}^{-1}_{ij} \phi_j.
\end{equation}
The above transformation can be achieved by a sequence of two-orbital rotations.
\begin{itemize}
    \item \textbf{Step 1}
    \begin{eqnarray*}
    &\phi_1 \leftarrow -\frac{1}{\sqrt{2}} \phi_1 + \frac{1}{\sqrt{2}} \phi_6, \quad \phi_1 \leftarrow \frac{1}{\sqrt{2}} \phi_1 + \frac{1}{\sqrt{2}} \phi_6
    \\
    &\phi_2 \leftarrow -\frac{1}{\sqrt{2}} \phi_2 + \frac{1}{\sqrt{2}} \phi_5, \quad \phi_5 \leftarrow \frac{1}{\sqrt{2}} \phi_2 + \frac{1}{\sqrt{2}} \phi_5
    \\
    &\phi_3 \leftarrow -\frac{1}{\sqrt{2}} \phi_3 + \frac{1}{\sqrt{2}} \phi_4, \quad \phi_4 \leftarrow \frac{1}{\sqrt{2}} \phi_3 + \frac{1}{\sqrt{2}} \phi_4
    \end{eqnarray*}
    \item \textbf{Step 2}
    \begin{eqnarray*}
    &\tilde{\alpha}_1 \leftarrow -\frac{1}{\sqrt{3}}\phi_1 + \sqrt{\frac{2}{3}} \phi_2, \quad \phi_2 \leftarrow \sqrt{\frac{2}{3}} \phi_1 + \frac{1}{\sqrt{3}}\phi_2
    \\
    &\phi_5 \leftarrow -\frac{1}{\sqrt{3}}\phi_5 + \sqrt{\frac{2}{3}} \phi_6, \quad \tilde{\alpha}_6 \leftarrow \sqrt{\frac{2}{3}} \phi_5 + \frac{1}{\sqrt{3}}\phi_6
    \end{eqnarray*}
    \item \textbf{Step 3}
    \begin{eqnarray*}
    \tilde{\alpha}_2 &\leftarrow -\frac{1}{\sqrt{2}} \phi_2 - \frac{1}{\sqrt{2}} \phi_4, \quad \tilde{\alpha}_4 \leftarrow -\frac{1}{\sqrt{2}} \phi_2 + \frac{1}{\sqrt{2}} \phi_4
    \\
    \tilde{\alpha}_3 &\leftarrow -\frac{1}{\sqrt{2}} \phi_3 + \frac{1}{\sqrt{2}} \phi_5, \quad \tilde{\alpha}_5 \leftarrow \frac{1}{\sqrt{2}} \phi_3 + \frac{1}{\sqrt{2}} \phi_5
    \end{eqnarray*}
\end{itemize}
We then arrive at the atomic-like orbitals $\tilde{\alpha}_i$'s.

\bibliography{qchem}

\end{document}